\documentclass[reprint, twocolumn, amsmath, amssymb, aps, pre, showpacs]{revtex4-1}
\usepackage[utf8]{inputenc}
\usepackage{graphicx}
\usepackage{color}
\begin{document}

\title{Grain coarsening in two-dimensional phase-field models with an orientation field}

\author{Bálint Korbuly$^1$, Tamás Pusztai$^1$, Hervé Henry$^2$, Mathis Plapp$^2$, Markus Apel$^3$, and László Gránásy$^{1,4}$}
\address{$^1$Institute for Solid State Physics and Optics, Wigner Research Centre for Physics, P.O. Box 49, H-1525 Budapest, Hungary}
\address{$^2$Laboratoire Physique de la Mati$\grave{e}$re Condens\' ee, \'Ecole Polytechnique, CNRS, Universit\'e Paris-Saclay, 91128 Palaiseau Cedex, France}
\address{$^3$Access e.V., Intzestr. 5, 52072 Aachen, Germany}
\address{$^4$BCAST, Brunel University, Uxbridge, Middlesex, UB8 3PH, United Kingdom}



\begin{abstract} 
In the literature, contradictory results have been published regarding the form of the limiting (long-time) grain size distribution (LGSD) that characterizes the late stage grain coarsening in two-dimensional and quasi-two-dimensional polycrystalline systems. While experiments and the phase-field crystal (PFC) model (a simple dynamical density functional theory) indicate a lognormal distribution, other works including theoretical studies based on  conventional phase-field simulations that rely on coarse grained fields, like the multi-phase-field (MPF) and orientation field (OF) models, yield significantly different distributions. In a recent work, we have shown that the coarse grained phase-field models (whether MPF or OF) yield very similar limiting size distributions that seem to differ from the theoretical predictions. Herein, we revisit this problem, and demonstrate in the case of OF models [by R. Kobayashi {\it et al.}, Physica D {\bf 140}, 141 (2000) and H. Henry {\it et al.} Phys. Rev. B {\bf 86}, 054117 (2012)] that an insufficient resolution of the small angle grain boundaries leads to a lognormal distribution close to those seen in the experiments and the molecular scale PFC simulations. Our work indicates, furthermore, that the LGSD is critically sensitive to the details of the evaluation process, and raises the possibility that the differences among the LGSD results from different sources may originate from differences in the detection of small angle grain boundaries.
\end{abstract}

\maketitle

\section{Introduction}

The majority of the solid matter we use appears in a polycrystalline form (including technical alloys, concrete, polymers, minerals, drugs, sugar, salt, cholesterol, peptides, etc.); i.e., they are composed of a large number of small crystallites. The properties of such materials depend on the size-, composition-, and shape-distributions of the crystallites they consist of. This also appears to be the case for polycrystalline thin metal films, where the properties of the grain size distribution may, e.g., influence the quality of metallization of semiconductors in electronics industry (by influencing the current-carrying capability via the resistance to electromigration \cite{ref1, ref2}). The distributions that characterize the polycrystalline structure can be influenced by the conditions of preparation (e.g., solidification, electrodeposition or sputtering) and subsequent processing, including heat treatments. As a result, the understanding of the grain coarsening process and the ability to predict the associated properties of the polycrystalline matter are of high technological importance, and have been the subject of intensive experimental and theoretical research. 

Grain coarsening is  mostly  due to the motion of grain boundaries and triple junctions  in a way that leads to the reduction of the excess free energy associated with the grain boundary network. While this process is fairly simple, the complexity of the geometry makes it difficult to predict the limiting grain size distribution (LGSD). Theory, experiments, and simulations agree that there exists a (time invariant) limiting grain size distribution, which evolves in a self-similar way, and the time dependence of the average grain size can be expressed as $\langle R \rangle = k t^n$, where $n$ is the growth exponent \cite{ref1, ref3}. Despite long-standing efforts, no convincing theoretical derivation of the LGSD has been proposed yet. Moreover, other phenomena such as grain rotation, elasticity, and anisotropy may affect the LGSD, adding to the difficulty of the task.

Herein, we concentrate on grain coarsening in two dimensional (2D) systems. The present results are expected to be relevant to 2D multi-grain structures, including thin metallic/ceramic films, colloidal aggregates, and plasma crystals. In the following paragraphs we briefly review the 2D results from experiment, theory, and simulation. This is a huge area indeed, and we can only summarize the main results here. Although the relevant process have been studied for some time, this brief review will show that the results are not without contradiction and that no clear understanding of the discrepancies among them has been reached yet.

\begin{figure}[tb]
  \includegraphics[width=0.99\linewidth]{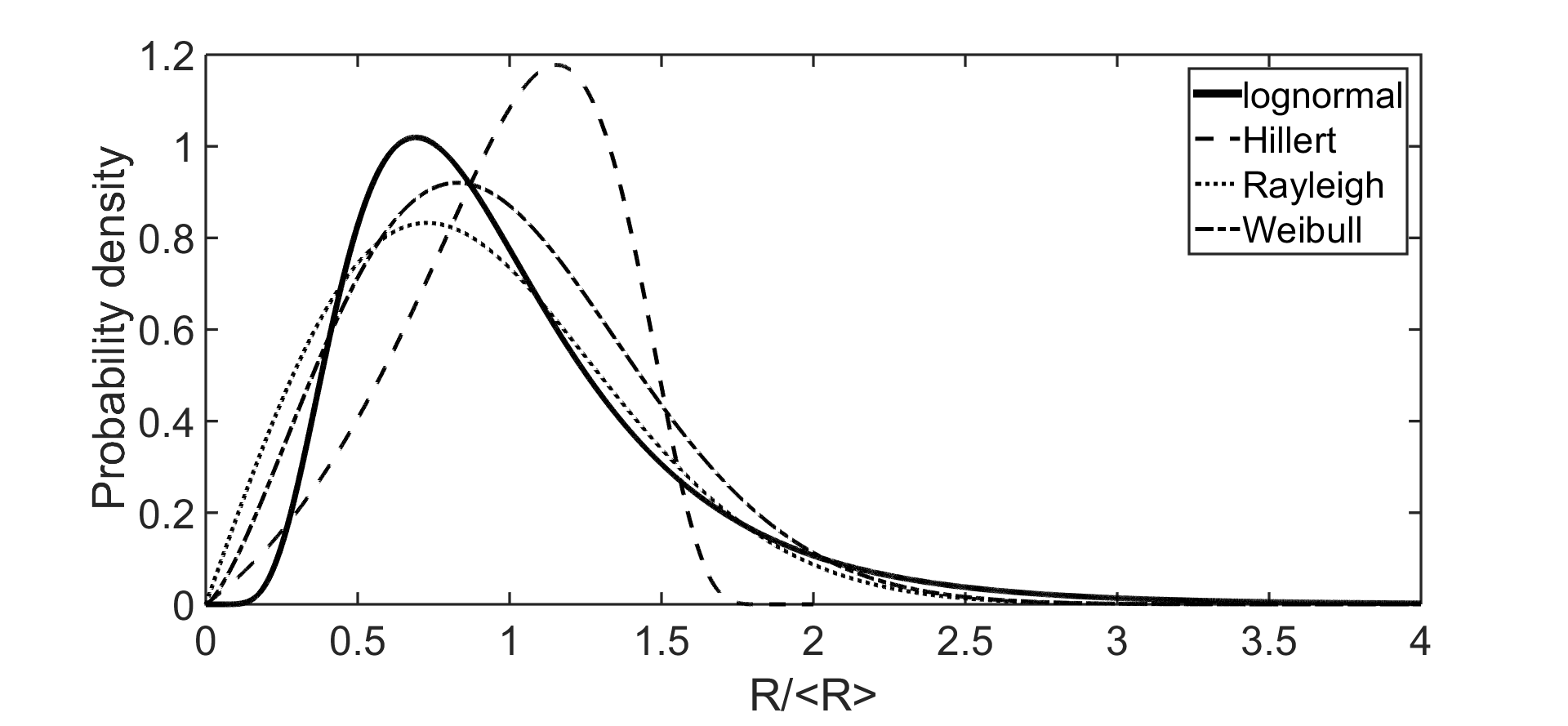}
  \caption{\label{fig:the_vs_exp} The limiting grain size distributions theoretically predicted for two dimensions. Hillert's distribution (dashed line) is free from adjustable parameters. Parameters of the other distributions were fitted to experimental results shown in Fig. A1 of Barmak {\it et al.} \cite{ref1}. Solid line: lognormal distribution ($\sigma = 0.5, \mu = -0.12$, \cite{ref1}); dotted line: Rayleigh distribution (proposed by Louat) with fitted parameters $\alpha = 1.030$ and $\beta = 2$; dash-dot line: Weibull distribution with fitted parameters $\alpha = 1.001$ and $\beta = 2.367$. The lognormal distribution gives a nearly exact representation of the experimental data for Al and Cu films (see Fig. A1 of Ref. \cite{ref1}). Here, $R$ is the equivalent radius computed from the grain area, whereas $\langle R \rangle$ is its arithmetic average over all grains.
}
\end{figure}

A recent detailed {\it experimental} study that summarizes results on 27 sputter deposited Al and Cu films \cite{ref1} indicates a lognormal LGSD with $\sigma = 0.5$ and $\mu = -0.12$ (Fig. \ref{fig:the_vs_exp}), and $n \approx 1/2$ for the growth exponent. The lognormal distribution, 
\begin{align}
p(x) = \frac{1}{x \sigma \sqrt{2 \pi}} \mathrm{exp}\bigg \{ -\frac{[ln(x) - \mu]^2}{2\sigma^2} \bigg \}, 
\end{align}
\noindent where $x$ is the linear size of the grains evaluated from their area, was found to be fairly robust. It remains an accurate description under a broad range of experimental conditions, including purity of the sputtering target, type of substrate, film thickness, temperature of deposition, actual and homologous annealing temperatures, the time of annealing, the grain size, and the twin density within the grains \cite{ref1}. This result is in agreement with some previous experimental results on thin films \cite{ref4,ref5}, while other works indicate deviations from the lognormal distribution  \cite{ref6}. It is also worth noting that significantly smaller $n$ values have also been reported in other works \cite{ref7,ref8}. The main cause for grain coarsening is the excess free energy of the grain boundary network, which relaxes via curvature driven migration of the grain boundaries \cite{ref9}. The velocity of the latter is often sought in the form $v = \kappa M$, where $\kappa$ is the curvature, and $M$ the grain boundary mobility. It is expected that other phenomena, such as grain rotation, elasticity, anisotropy, etc. may also contribute to the dynamics of grain coarsening \cite{ref10}. 

There appears to be no consensus regarding the {\it theoretical} form of LGSD in 2D: Kolmogorov \cite{ref11} has shown that starting from a single grain of known volume, repeated crushing will lead to a lognormal distribution of particle volumes. In Feltham's early work \cite{ref4} a lognormal distribution was postulated. In a more sophisticated theoretical treatment, that follows  the route of Lifshitz, Slyozov, and Wagner used in addressing Ostwald ripening \cite{ref12,ref13}, Hillert \cite{ref14} derived the distribution function for two dimensional isotropic growth in multigrain structures 
\begin{align}
p(x) = (2e)^2 \frac{2x}{(2-x)^4} \mathrm{exp} \bigg \{ - \frac{4}{2 - x} \bigg \}.
\end{align}
\noindent Regarding grain growth as a statistical phenomenon taking place via the growth of faces, Louat \cite{ref15} derived a Rayleigh type grain size distribution. A statistical extension of Feltham's approach by Kurtz and Carpay, in turn, supports the lognormal form \cite{ref16}. The empirical Weibull distribution, corresponding to 
\begin{align}
p(x) = \frac{\beta}{\alpha}\left( \frac{x}{\alpha} \right)^{(\beta-1)} \mathrm{exp} \bigg \{ - \left( \frac{x}{\alpha} \right) ^{\beta} \bigg \}, 
\end{align}
\noindent and its special case the Rayleigh distribution $(\beta = 2)$  were also used to approximate the experimentally observed LGSDs \cite{ref17}. While the use of Weibull distribution is empirical in this context, the Rayleigh distribution was derived by Louat \cite{ref15}. Relying on the Neumann-Mullins growth law, in more recent works Pande and Cooper \cite{ref18,ref19} deduce a Fokker-Planck equation for the grain size distribution, which yields a self-similar asymptotic solution that can be reached from arbitrary initial state. They propose an approximate analytical solution for LGSD, that can be tuned by varying a single parameter between the Rayleigh distribution (where all curvature effects are neglected) and Hillert's model (where the drift velocity due to curvature is the only driving force) \cite{ref19}. Seeking a flexible empirical LGSD distribution, Rickman {\it et al.} quantify the deviation from the lognormal distribution in terms of a cumulant expansion tailored to the lognormal distribution \cite{ref20}.   

The models discussed above predict $n = 1/2$. A comparison of the predicted LGSDs with parameters taken from fitting to recent experimental data of \cite{ref1} are presented in Fig. \ref{fig:the_vs_exp}. In agreement with previous work \cite{ref1,ref5,ref19}, these distributions appear to be less satisfactory than the lognormal form. 
 
\begin{figure}[t]
  \includegraphics[width=0.99\linewidth]{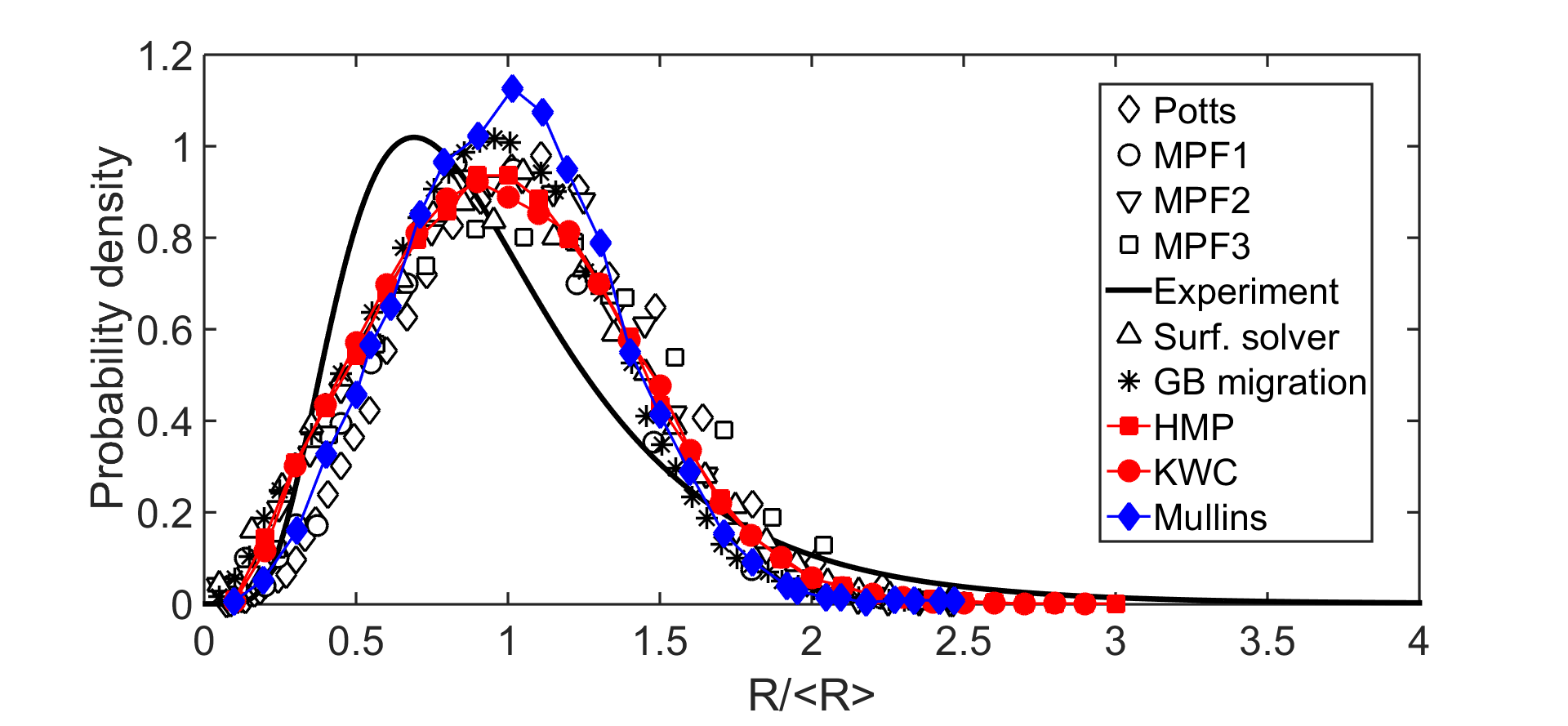}
  \caption{\label{fig:simcomp} (color online) The limiting grain size distributions from computer simulations in two dimensions: diamonds -- Potts model \cite{ref22}; circles, triangles pointing downward, and squares --  multi-phase-field models \cite{ref22, ref23, ref24}, respectively; solid line -- experiment Fig. A1 from Ref. \cite{ref1}; upward pointing triangles -- surface solver \cite{ref28}; stars -- grain boundary migration model by Moldovan {\it at al.} \cite{ref10}; full squares and circles -- orientation field models from Refs. \cite{ref25, ref26}, respectively. Note the similarity of the LGSDs from different types of simulations, implying a Mullins-type \cite{ref9} generic numerical solution (full diamonds, taken from Ref. \cite{ref1}).}
\end{figure}

\begin{figure}[b]
  \includegraphics[width=0.99\linewidth]{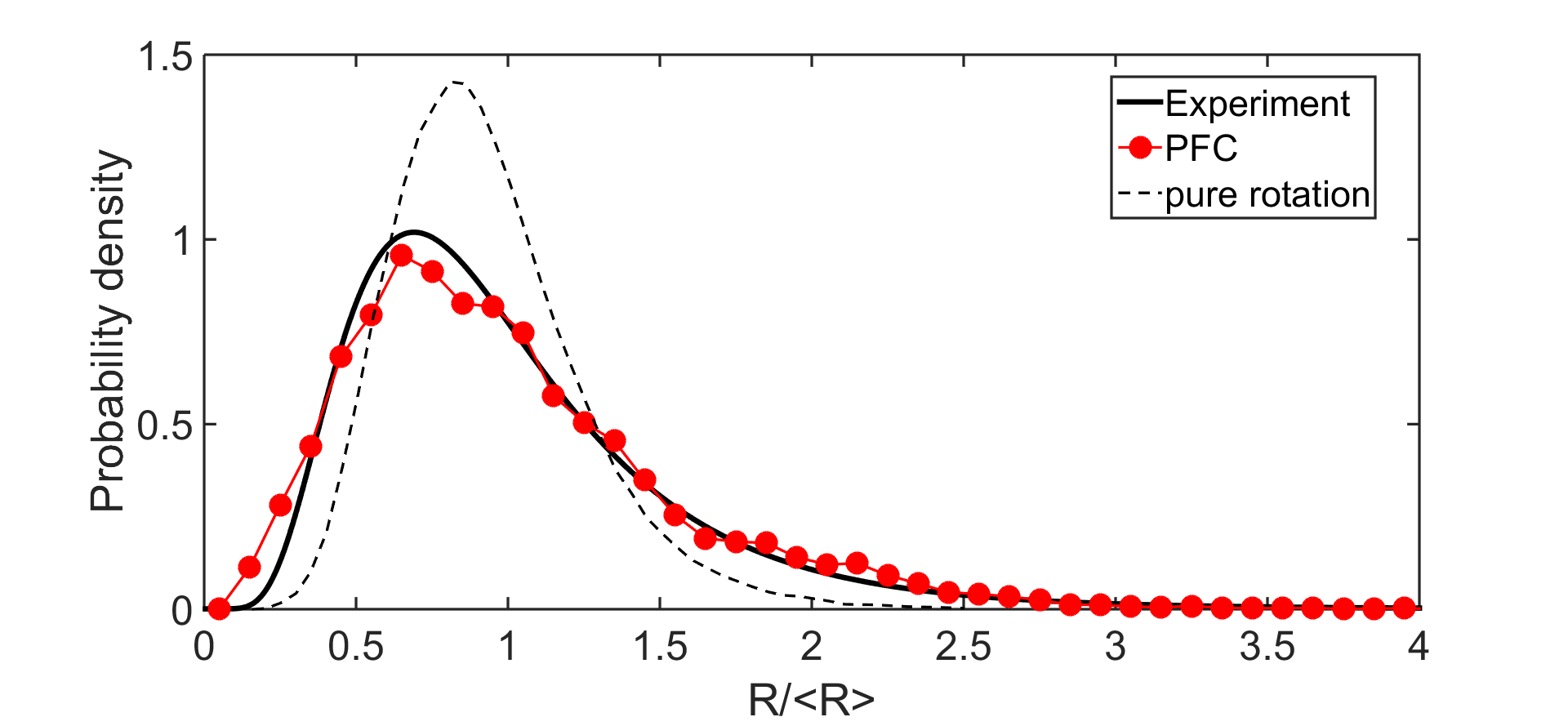}
  \caption{\label{fig:pfc} (color online) The limiting grain size distributions from computer simulations in two dimensions: solid line -- experiment \cite{ref1}; full circles -- phase-field crystal (PFC) simulation \cite{ref29}; grain coarsening via pure rotation \cite{ref10}. Note the excellent agreement between the PFC predictions and the experiments.}
\end{figure}

{\it Computer simulations} have also been used extensively to investigate grain coarsening. The results appear again somewhat contradictory. For example early results on Monte Carlo simulations for the Q = 32 2D Potts model \cite{ref21} indicate a fair agreement with Hillert's mean field model, yet a later Q = 72 Potts study reports a substantially different LGSD \cite{ref22}. The other 2D models predict rather similar limiting grain size distributions. This includes a grain boundary migration based numerical approach by Moldovan {\it et al.} \cite{ref10}, three different versions of the multi-phase-field theory (MPF1 \cite{ref22}, MPF2 \cite{ref23}, MPF3 \cite{ref24}), and two orientation field based models: one by Kobayashi, Warren, and Carter (KWC) \cite{ref25} and another by Henry, Mellenthin, and Plapp (HMP) \cite{ref26}. LGSDs for the OF models were reported in Ref. \cite{ref27}. Another approach that led to similar results employed a numerical surface solver to relax the interface energy \cite{ref28}. The predicted distributions are fairly similar (with some scattering), although different models and methods are compared. This finding might originate from the fact that these models all tend to reduce the free energy associated with the grain boundary network. Apparently, LGSD from these 2D simulations fit reasonably well to the Weibull type probability density. Although the investigated 2D simulations are {\it consistent with each other} (Fig. \ref{fig:simcomp}), and follow the behavior obtained using the model of Mullins \cite{ref9} in Ref. \cite{ref1}, yet they significantly differ from the lognormal distribution representing recent experimental data \cite{ref1}. These findings are consistent with earlier results of Pande and Cooper \cite{ref18,ref19}. Interestingly, simulations performed using a dramatically different process for coarsening, i.e. grain rotation, yield a lognormal distribution \cite{ref10}, yet much different from experiments (see Fig. \ref{fig:pfc}).

Remarkably, the only simulation results for LGSD that agree with the experiments are from the phase-field crystal (PFC) model, an approach that works on the molecular scale \cite{ref29} (see Fig. \ref{fig:pfc}). The PFC model developed by Elder {\it et al.} \cite{ref30} can be regarded as a simple dynamical density functional theory that incorporates the crystal structure, anisotropies, elasticity, and dislocations automatically (for a review on PFC see Ref. \cite{ref31}). As such, it contains a much richer physics that the previous models, including automatically elasticity, dislocation dynamics, grain rotation, molecular scale description of grain boundaries, etc. A possibility is that this richness of phenomena is responsible for the accurate LGSD it predicts. Herein, we raise a different possibility, which might explain at least partly the observed differences between LGSDs from simulations and experiments.

We use the orientation field models HMP and KWC to demonstrate that the limiting grain size distribution is {\it critically sensitive} to details of the evaluation of the number of the grains, especially to the resolution of the small angle grain boundaries. Variation of the misorientation, below which a grain boundary is not detected any more, yields a continuous transition between the lognormal distribution that can be observed for poorly resolved low angle grain boundaries (only a fraction of them are found), and the general behavior from 2D simulations shown in Fig. \ref{fig:simcomp}, obtained when the low angle grain boundaries are well resolved (the majority of them was found). This finding raises the possibility that improving the resolution of low angle grain boundaries in the experiments (and maybe in PFC simulations) might yield LGSDs falling closer to the results provided by the 2D simulations.   

The present paper is structured as follows: In Section II, we briefly recapitulate the essence of the orientation field theories HMP and KWC, while Section III specifies the materials properties and other conditions used in the simulations. Section IV describes the details of the evaluation methods employed for determining the LGSD, with a detailed analysis of different factors that influence the limiting distribution, and discuss the consequences. Finally, in Section V, we summarize the main results and offer a few concluding remarks.

\section{Orientation field models}
We employ here two phase-field models that describe polycrystalline solidification and grain coarsening on equal footing: the KWC and HMP models described in Refs. \cite{ref25} and \cite{ref26}, respectively. In these models the transition between the liquid and crystalline phases is monitored by a structural order parameter, the phase field $\phi(\mathbf{r}, t)$, whereas the local crystallographic orientation is specified by a scalar orientation field $\theta(\mathbf{r}, t)$, normalized so that $\theta \in  [0, 1]$, considering the crystal symmetry. For $k$-fold symmetry, the orientation angle $\vartheta$ can vary between 0 and $2 \pi/k$. Owing to the $k$-fold symmetry, $\vartheta$ angles outside this region are equivalent to a specific orientation angle inside the $[0, 2 \pi/k]$ regime. (Then $\theta = \vartheta/(2\pi/k) \in  [0, 1]$ describes all orientations.) Being an angular variable $\theta = 0$ and $\theta = 1$ are equivalent, and the magnitude of the orientation field difference is limited: $ |\Delta \theta| \le \frac{1}{2}$. This should be considered, e.g., when evaluating the differential operators acting on the orientation field. For the sake of simplicity, isotropic systems will be considered, yet the grain boundary energy depends naturally on the misorientation in these models, as will be displayed in Fig. \ref{fig:GB}(a) below.

\subsection{Free energy functional}  
 
The free energy of the respective system is a functional of the fields $\phi(\mathbf{r}, t)$ and $\theta(\mathbf{r}, t)$, which can be written in the following form for both orientation field approaches we use here: 
\begin{align}
F = \int d\mathbf{r} \bigg \{ \frac{\epsilon^2 T}{2}\left(\nabla\phi\right)^{2} &+ W T g(\phi) + \nonumber \\ 
&+ f_\mathrm{bulk}(\phi) + f_\mathrm{ori}(\phi,\nabla\theta) \bigg \}, 
\label{eq:freeenergy}
\end{align}
where the model parameters $\epsilon^2$ and $W$ are expressible in terms of the  free energy and thickness of equilibrium solid-liquid interface, and $T$ is the temperature. $f_\mathrm{bulk}$ switches between the free energy densities of the bulk solid and liquid phases according to the interpolating function $p(\phi)$ as follows
\begin{align}
f_\mathrm{bulk}(\phi) &= p(\phi) f_\mathrm{s}(T) + (1-p(\phi)) f_\mathrm{l}(T)\\
p(\phi) &= \phi^3 (10 - 15\phi + 6\phi^2),
\end{align}
where the free energy densities of the bulk solid and liquid phases, $f_\mathrm{s}(T)$ and $f_\mathrm{l}(T)$, respectively, can be taken from thermodynamic databases.

The two orientation field based models we use differ in the form of the orientational contribution to the free energy, $f_\mathrm{ori}(\phi,\nabla\theta)$: 
\begin{align}
f_\mathrm{ori}^\mathrm{KWC}(\phi,\nabla\theta) &= r(\phi) T \left\{ H_{1} |\nabla\theta|+H_{2} (\nabla\theta)^{2}\right\} \\
f_\mathrm{ori}^\mathrm{HMP}(\phi,\nabla\theta) &= q(\phi) T H (\nabla\theta)^{2}
\label{eq:fori}
\end{align}
where $r(\phi) = \phi^4$ and  $q(\phi) = \frac{7\phi^{3}-6\phi^{4}}{(1-\phi)^{3}}$. These functions differ slightly form the original formulations \cite{ref32}. Parameters $H$, $H_1$, and $H_2$ tune the strength of the respective gradient terms, and scale the grain boundary energy.

\subsection{Equations of motion (EOMs)}

The time evolution of the system is assumed to follow standard variational dynamics of non-conserved fields:
\begin{align}
\frac{\partial\phi}{\partial t} &= -M_{\phi} \frac{\delta F}{\delta\phi} + \zeta_\phi\
\label{eq:EOM1}
\end {align}
\begin{align}
\frac{\partial\theta}{\partial t} &= -M_{\theta} \frac{\delta F}{\delta\theta} + \zeta_\theta ,
\label{eq:EOM2}
\end{align}
where $M_{\phi} (= const.)$ and $M_{\theta}$ are the respective mobilities, $M_{\theta} = M_{\theta,S} + (M_{\theta,L} + M_{\theta, S}) p(\phi)$, whereas $\zeta_{\phi}$ and $\zeta_{\theta}$ are noise terms that represent the fluctuations of the respective fields. Their correlators are $\langle \zeta_i(\mathbf{r}, t),\zeta_i(\mathbf{r}', t') \rangle = \omega_{ij} 2kTM_{ij} \delta(\mathbf{r} - \mathbf{r}') \delta(t - t')$, where $\omega_{ij}$ is the noise strength coefficient for the $i$th field in the $j$th phase ($i = \phi, \theta$ and $j = L,S$). For further details of EOMs and their dimensionless forms see Appendix I. 

We found that in the orientation field models topological defects appear that resemble the disclinations observed in 2D atomistic models using the hexatic order parameter \cite{ref33}, which may influence the motion of grain boundaries and trijunctions. This phenomenon has already been detected by Warren {\it et al.} \cite{ref25}(c). The properties of these defects and possible ways to remove them via a complex orientation field, a 3-component orientation field, and other means will be addressed elsewhere \cite{ref34}. Herein, we use $\omega_{\phi,S,L} = 0,$ $\omega_{\theta,S} = 0$ and $\omega_{\theta,L} = 0.1$, to remove the pinning effect of these topological defects in the orientation field.

\subsection{Numerics}

The dimensionless form of Eqs. (\ref{eq:EOM1}) and (\ref{eq:EOM2}) were solved numerically on rectangular grids of different sizes, using finite difference discretization combined with  explicit forward Euler time stepping, while prescribing periodic boundary conditions. Parallel codes were developed for a CPU cluster and GPU cards. The computations were performed on  GPU cards of various types.

\section{Materials properties and other conditions}

In the simulations, we used the physical properties of pure Ni. The volumetric free energy difference between the liquid and the solid was estimated using Turnbull's linear approximation: $\Delta f = \Delta H_f (T_f - T)/T_f$ \cite{ref35}, where $\Delta H_f = 2.61 \times 10^9$ J/m$^3$ and $T_f = 1728$ K are the volumetric heat of fusion and the melting point, respectively. The thickness of the equilibrium solid-liquid interface were taken as $d = 2$ nm, which is of the order of magnitude of results from molecular dynamics simulations \cite{ref36}. The free energy of the solid-liquid interface, $\gamma_{SL} = 0.364$ J/m$^2$ was taken from the compilation \cite{ref37}; whereas a molar volume $V_m = 6.59 \times 10^{-6}$ m$^3$/mol was employed. Isothermal computations were performed at $T = 974$ K. The diffusion coefficient and the characteristic length applied in making the EOMs dimensionless (see Appendix I) were $D_L = 10^{-9}$ m$^2$/s and $\xi = 40$ nm, respectively. If not stated otherwise, the dimensionless spatial and time steps were chosen as $\Delta \tilde{x} = 3.125 \times 10^{-3}$ and $\Delta \tilde{t} = 1.4844 \times 10^{-8}$, which ensured the numerical stability of the solution. The dimensionless phase-field mobility was taken as $\tilde{M}_{\phi}^{HMP} = \tilde{M}_{\phi}^{KWC} = 0.9$.

\begin{figure}[t]
 \includegraphics[width=0.49\linewidth]{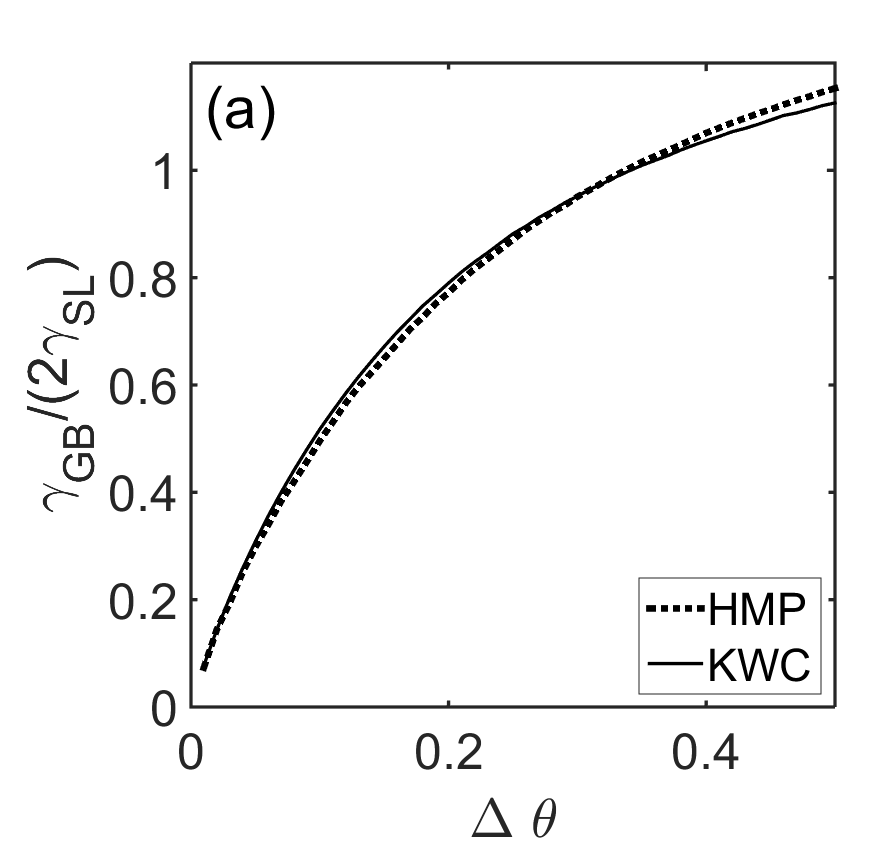}  
 \includegraphics[width=0.49\linewidth]{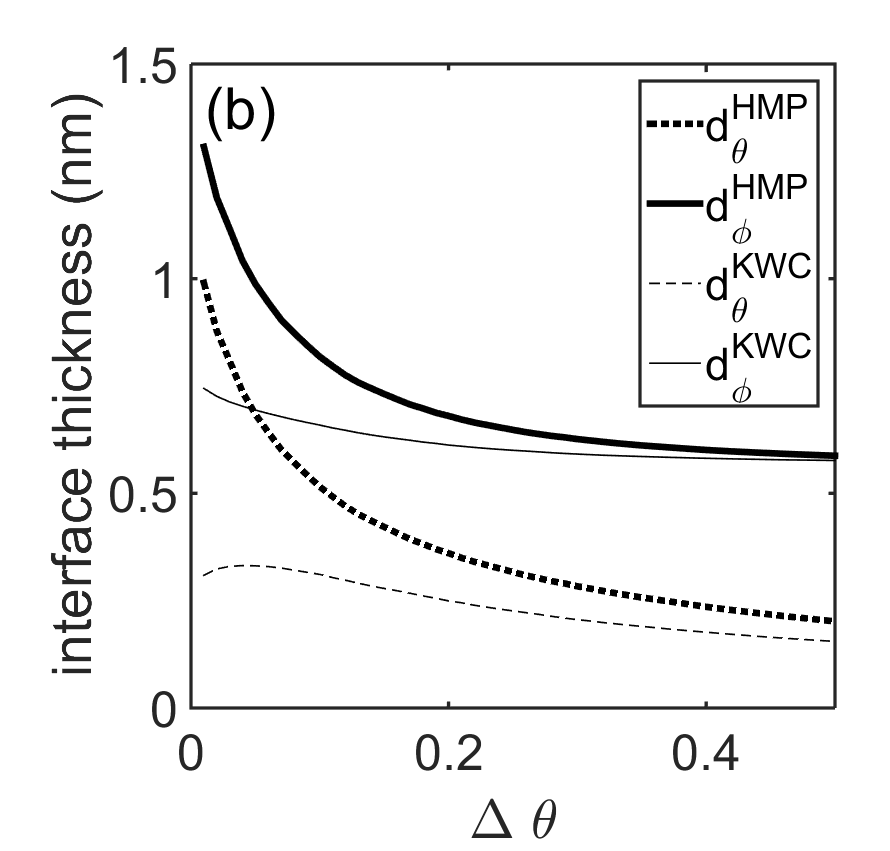} \\  
 \includegraphics[width=0.49\linewidth]{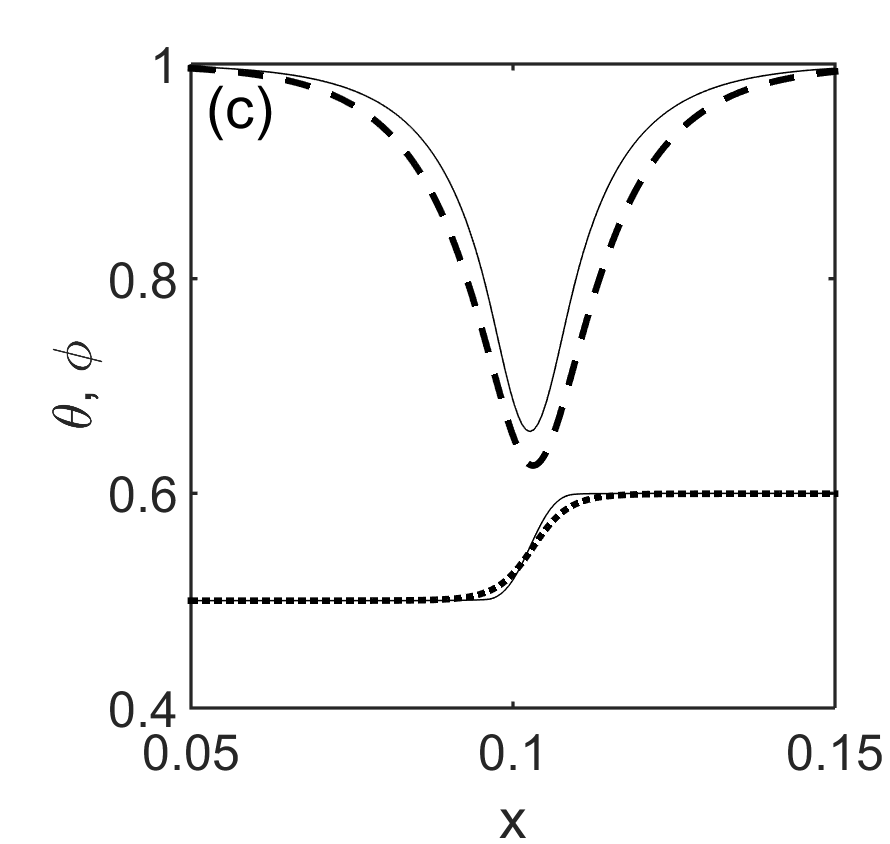}  
 \includegraphics[width=0.49\linewidth]{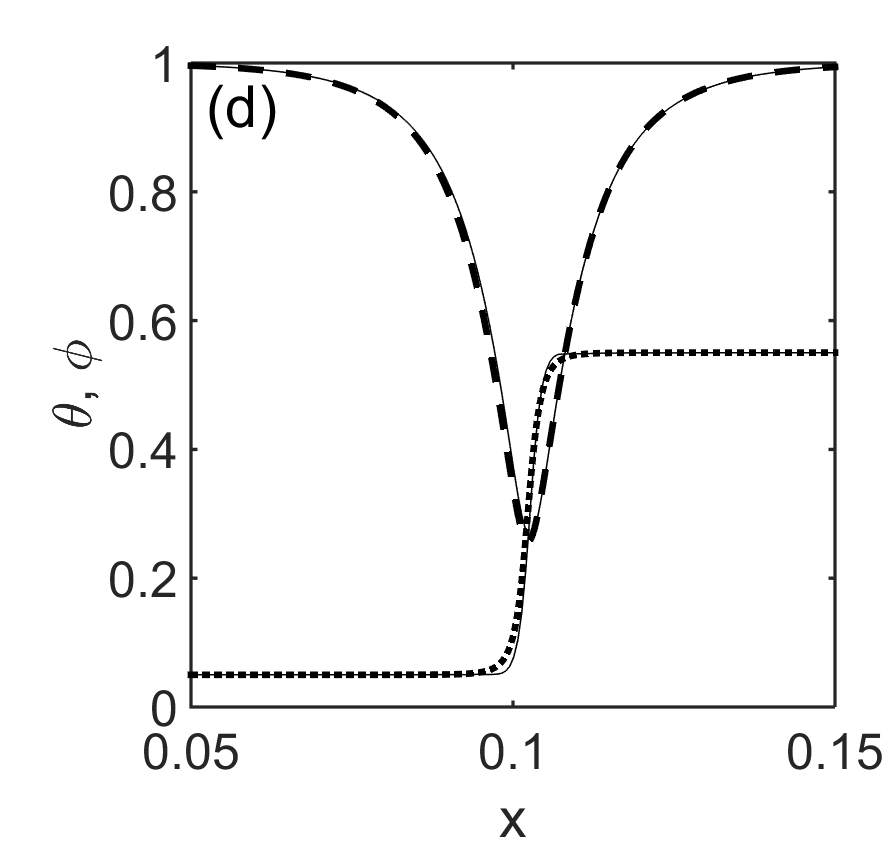}\\
  \caption{\label{fig:GB} Properties of the grain boundaries predicted by the orientation field models (the heavy lines are for the HMP model, whereas the thin lines are for the KWC): (a) grain-boundary energy vs. misorientation; (b) interface thickness vs. misorientation (solid lines: phase field, dotted lines: orientation field, $d_{\phi}$ denotes the half-width of the depression, whereas $d_{\theta}$ stands for the $10\% - 90\%$ interface thickness); (c) and (d) cross-interfacial phase- and orientation field profiles at $\Delta \theta = 0.1$ and 0.5, respectively. $x$ is a dimensionless length.}
\end{figure}

\subsection{Synchronizing the HMP and KWC models}

Herein, we synchronize the HMP and KWC models so that they display similar properties (grain-boundary energy and time scale for grain-rotation) under the same circumstances; i.e., they can be viewed as models describing polycrystalline solidification/grain coarsening in the same matter.

To match the properties of the solid-solid interface, we have first chosen the dimensionless parameters (see Appendix) $\alpha_2= 1.0$, $\tilde{M}_{\theta, S}^{HMP} = 28.8$, and $\tilde{M}_{\theta,L}^{HMP} = 92.68$, and then varied the parameters of the KWC model until similar grain boundary energies, interface profiles, and time scales were obtained. This is realized by the following set of dimensionless parameters $\alpha_1 = 10.64$, $\alpha_2 = 30.0$,  $\tilde{M}_{\theta, S}^{KWC} = 7.2 \times 10^{-4}$, and $\tilde{M}_{\theta,L}^{KWC} = 1080$. These data ensure that in the bulk crystal ($\phi = 1$) the orientation mobility is negligible in both cases. 

\begin{figure}[t]
 \includegraphics[width=0.99\linewidth]{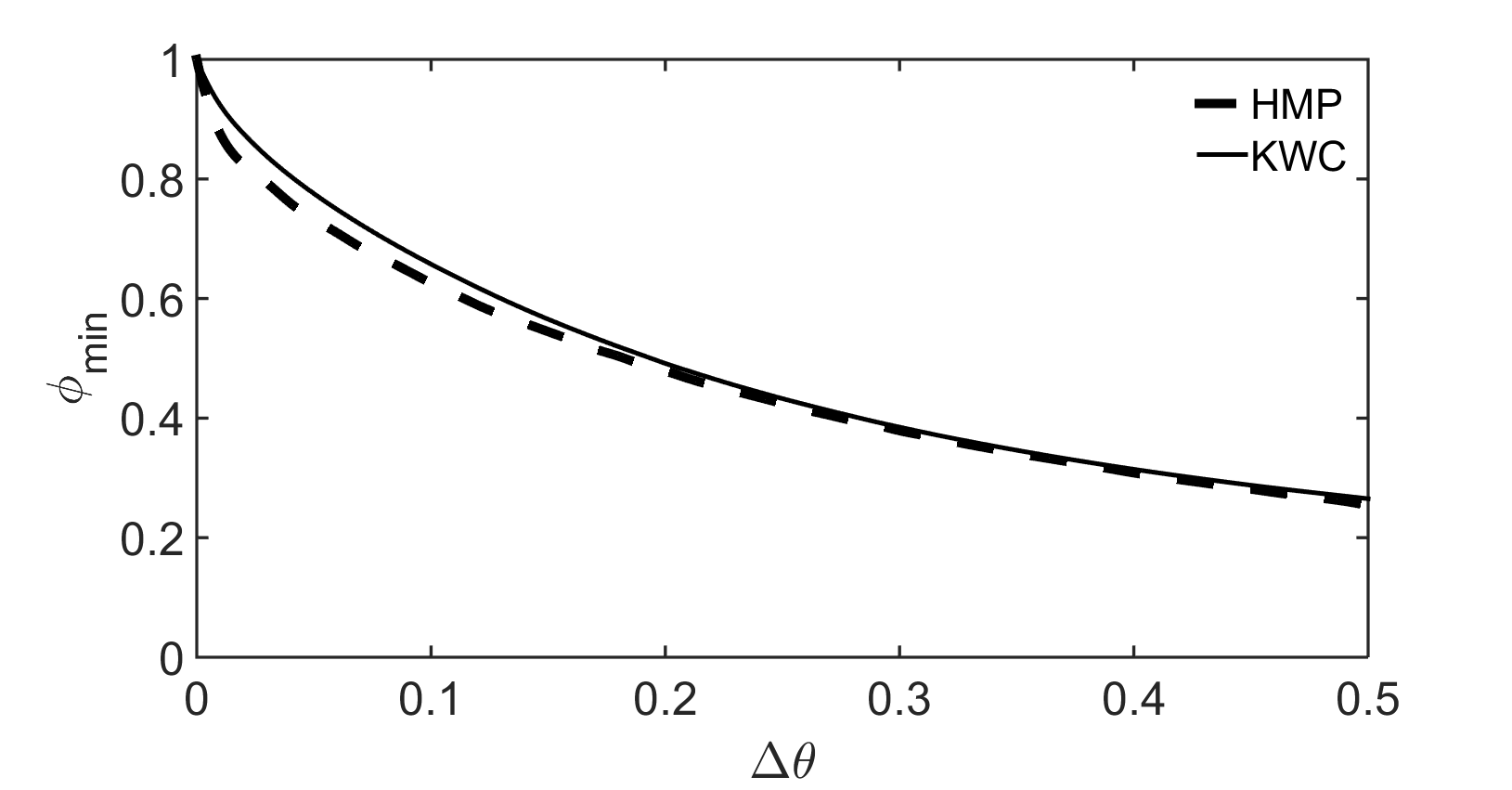} \\  
  \caption{\label{fig:phdepth} Phase-field minimum at the grain boundary vs. misorientation for the HMP (heavy dashed line) and KWC (light continuous line) models.}
\end{figure}

The respective grain boundary energies and interfacial profiles are shown in Fig. \ref{fig:GB}. The two models provide indeed fairly coherent predictions. One of the important features is the depression of the phase field at the grain boundary, which indicates that the crystalline order is disturbed within the grain boundary. For high misorientations the phase field may get as low as $\phi \sim 0.3$ here, but may reach $\phi = 0$ at the melting point that implies the presence of a highly disordered structure at the grain boundary, a behavior similar to the one observed in molecular dynamics simulations \cite{ref38}. The relationships between the depth of the phase-field depression and the misorientation are shown for the two models in Fig. \ref{fig:phdepth}.

\begin{figure}[t]
 \includegraphics[width=0.99\linewidth]{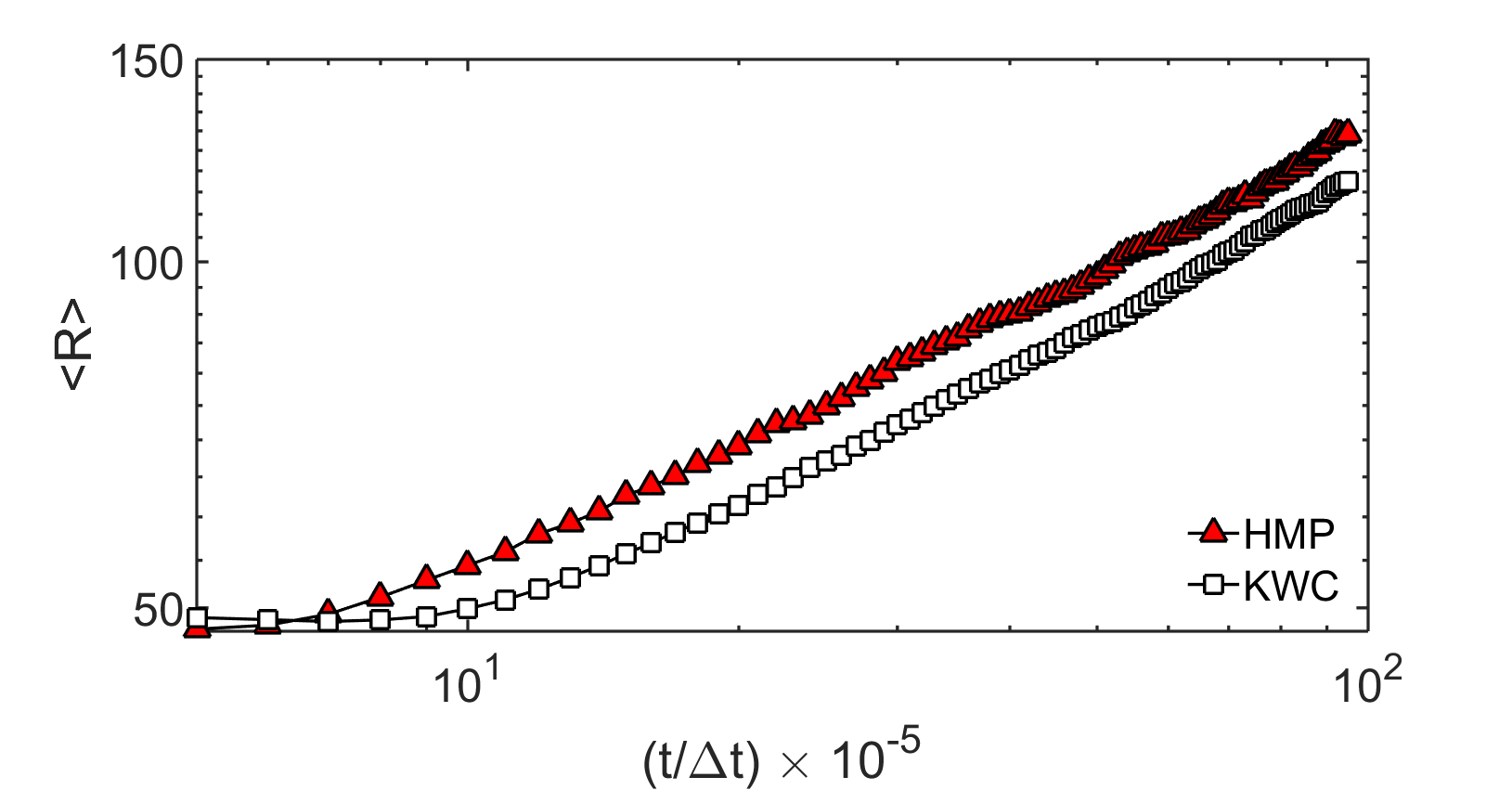} \\  
  \caption{\label{fig:dynmatch} (color online) Grain growth predicted by the two models are compared. With the present choice of parameters the growth rates of the HMP and KWC models are of comparable magnitude.}
\end{figure}

Next, we compare grain growth in the models. With the present choice of mobilities, the growth rates of the two models are of comparable magnitude [see Fig. \ref{fig:dynmatch}]. The simulation data were fitted by the formula $\langle R \rangle = A  (t - t_0)^n$, where $t_0$ is the hypothetical starting point of growth corresponding to $\langle R \rangle = 0$.  The fitting was performed so that only data beyond the end of an apparent initial transient period  ($8 \times 10^5$  and $10^6$ time steps for the HMP and KWC models, respectively) were considered. This procedure yielded growth exponents comparable to those from other models (HMP: $n=0.45\pm0.01$; KWC: $n=0.51\pm0.01$).

\subsection{Evaluation of LGSD}

In determining the LGSD, we used typically eight $4096^2$ simulations performed with different initializations of the random number generator. We used one of the following methods to generate the initial grain size distribution: (i) added noise to the equation of motion of the phase field to initiate nucleation and growth; (ii) we placed randomly oriented particles of 30 pixel radius randomly in the simulation box and let them grow; (iii) we also explored the case when the randomly oriented particles were placed on a square grid. The respective simulations yielded similar LGSDs, indicating that the long time behavior is not sensitive to the initial conditions. 

Unless stated otherwise, the grains were identified by the watershed algorithm of MATLAB \cite{ref39}. The watershed algorithm finds  "catchment basins" and "watershed ridge lines" in an image by treating it as a surface, where light pixels represent high elevations and dark pixels represent low elevations. It has been applied to evaluate the grain size distribution from the $(1 -  \phi)$ maps. In this case, the catchment basins correspond to the grains and the watershed ridge lines are the grain boundaries.

We have found that even merging eight $4096^2$ simulations (that contained initially $\sim 17,200$ randomly oriented grains) the grain size histograms show some visible scattering that makes difficult the comparison of LGSDs not very far from each other. It was, however, detected that at late stages of the evolution of the grain boundary network (in which we are interested anyway), fitting an appropriate analytic formula to the histograms (lognormal or Weibull), the results are essentially indistinguishable (i.e., they match to several states of the grain boundary network with a similar accuracy). In the case of such states of the system, it makes sense to merge the reduced distributions, to reduce the scattering of the histograms. One needs to be careful, however, since this is possible only at the end of the relaxation process, where the limiting size distribution is established. We have employed this technique for the evaluation of all LGSDs shown. For the details see Appendix II.

We have also evaluated the probability distribution of the misorientations along the grain boundaries identified by the watershed algorithm. For the details see Appendix III.

\begin{figure}[t]
 (a) \includegraphics[width=0.425\linewidth]{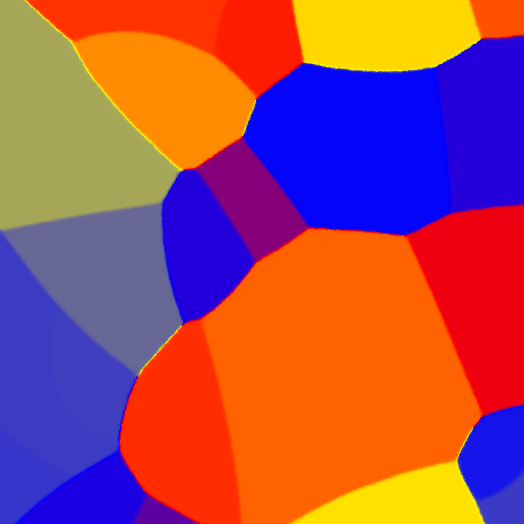}  
 (b) \includegraphics[width=0.425\linewidth]{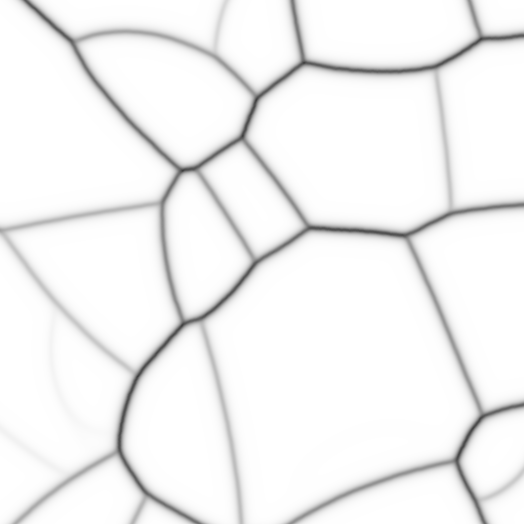} \\  
 (c) \includegraphics[width=0.425\linewidth]{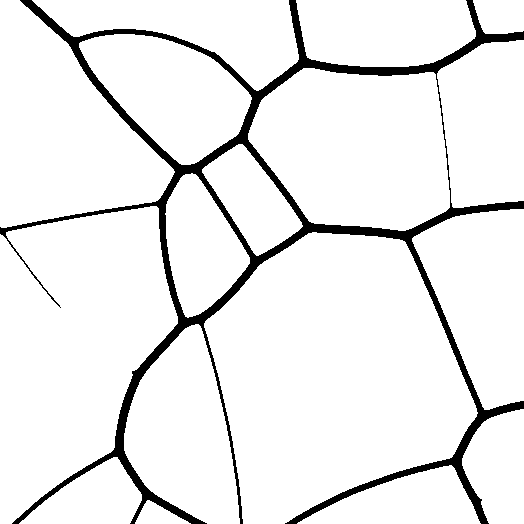}  
 (d) \includegraphics[width=0.425\linewidth]{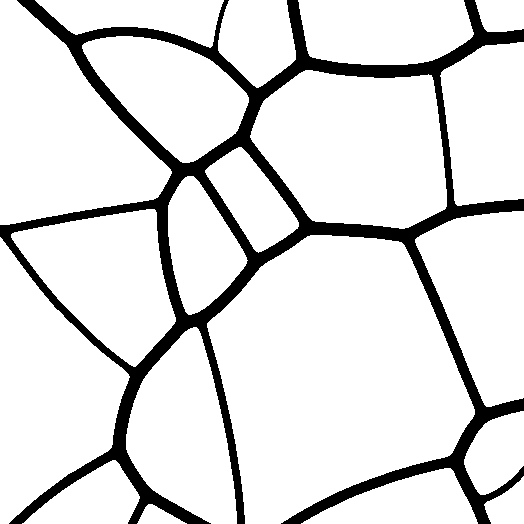}\\
 (e) \includegraphics[width=0.425\linewidth]{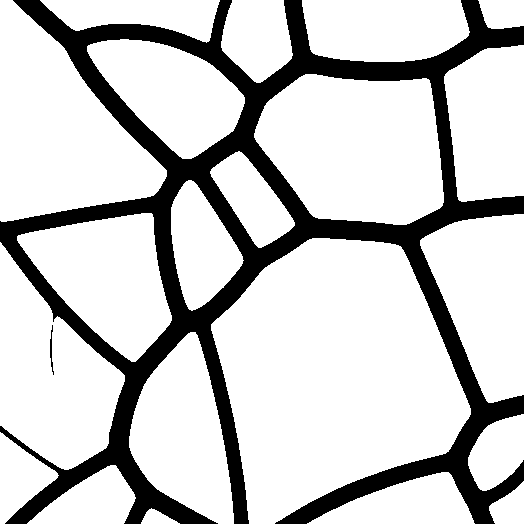}  
 (f) \includegraphics[width=0.425\linewidth]{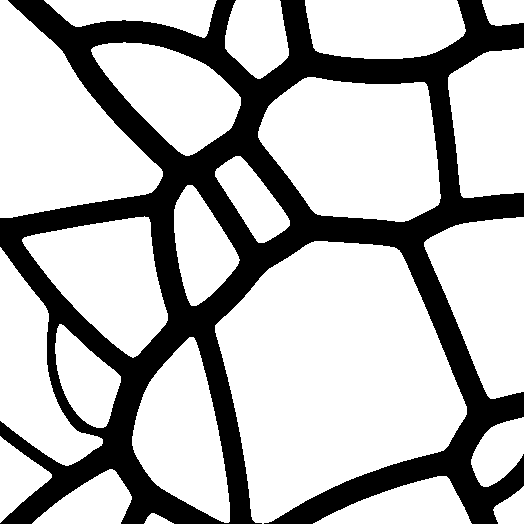}\\
  
\caption{\label{fig:eval} (color online) Finding the grains on the basis of the phase field depression at the grain boundaries in the HMP model (only a small fraction of a large simulation is shown). Original (a) orientation and (b) the gray-scale version of the phase field. (c)--(f) Black and white maps obtained by applying thresholds $\phi_{th} = 0.79, 0.89, 0.98,$ and $0.99$. If $\phi \le \phi_{th}$ the pixel is painted black, the rest is colored white. Note that recognition of the small angle grain boundaries happens to different degrees in these images. Accordingly, the respective grain counts and the corresponding size distributions differ.}
\end{figure}

\begin{figure}[b]
 (a) \includegraphics[width=0.93\linewidth]{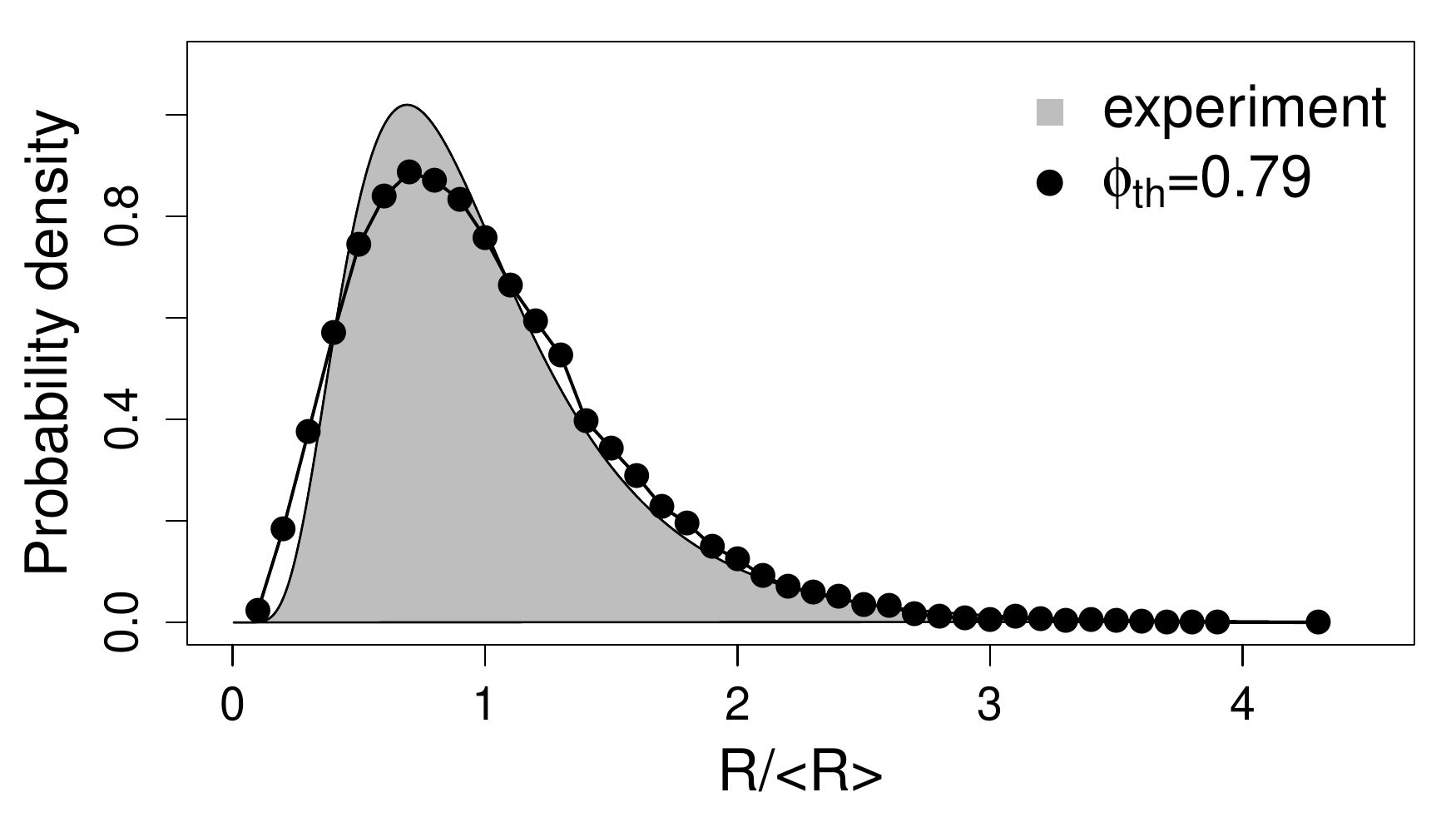}\\  
 (b) \includegraphics[width=0.93\linewidth]{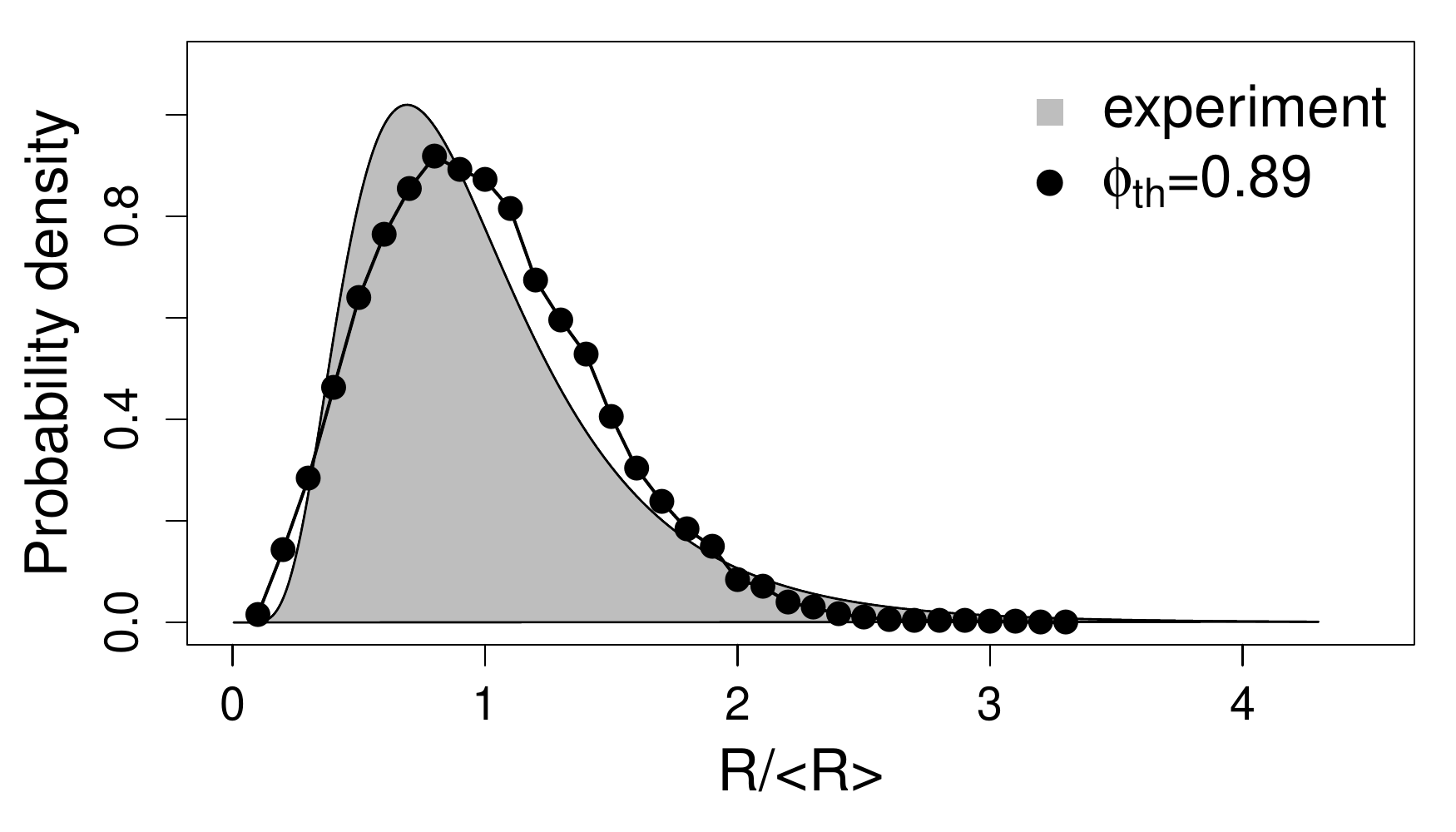}\\   
 (c) \includegraphics[width=0.93\linewidth]{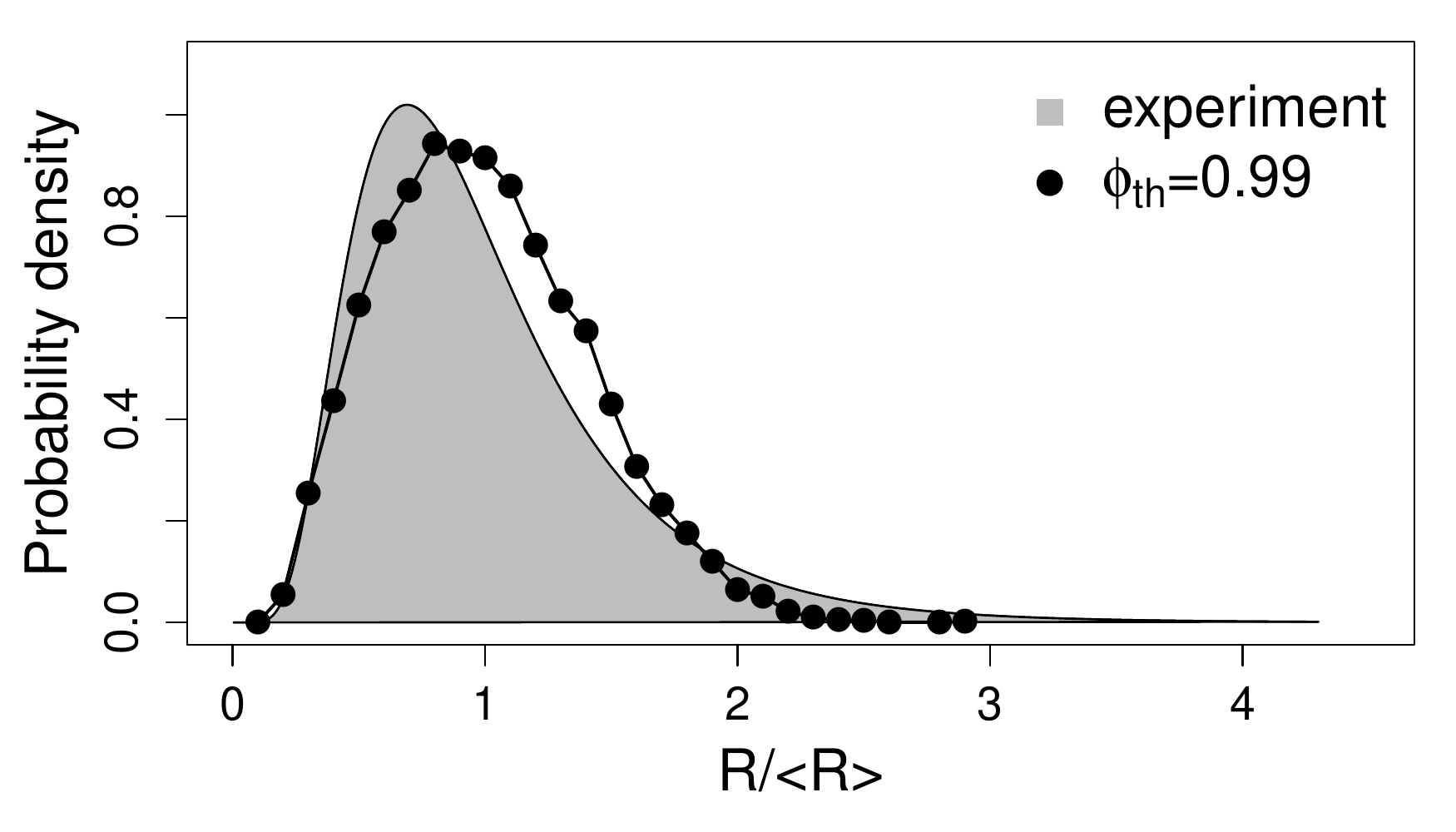}\\  
  
\caption{\label{fig:LGSDcomp} Limiting grain size distributions for the evaluations of the HMP simulation shown in Figs. \ref{fig:eval}(c), \ref{fig:eval}(d), and \ref{fig:eval}(f), obtained using the following threshold values: (a) $\phi_{th} = 0.79$, (b) $0.89,$ and (c) $0.99$. Note the similarity between the experimental distribution and the one obtained with the lowest value of the threshold $\phi_{th}$. Comparable results were obtained with the KWC model.}
\end{figure}

\section{Results and discussion}

The aim of the present investigation is to clarify whether differences in the accuracy of the evaluations can significantly influence the results for the LGSD. This was motivated by the fact that seemingly small changes in our evaluation procedures for the LGSD could yield significantly different distributions starting from the same raw data. Here, we pinpoint the reasons for this behavior and show that it is generic for orientation-field models. More precisely, the shape of the LGSD critically depends on the detection of small-angle grain boundaries. Beyond the models treated here, this observation raises several questions that will be discussed below.

\subsection{Methods for distinguishing the grains}

There are different possibilities to evaluate the grain size distribution from the phase- and orientation fields. Since the orientation field varies continuously across the grain boundaries and thus the same orientations may occur in a grain and at grain boundaries elsewhere, the recognition of grains is more complicated from the orientation field. Thus, in a previous work \cite{ref27}, we have opted for the identification of the grains via the depression of the phase field at the grain boundary, and used the watershed algorithm to locate them, yielding the red curves in Fig. \ref{fig:simcomp}. Herein we introduce a different approach that is also based on the phase-field map, yet relies on a threshold value in recognizing where the grain boundaries lie, allowing thus for a continuous tuning of the fraction of small angle grain boundaries considered in the evaluation. 

\subsection{Tunable approach to distinguish grains}

 As demonstrated by Figs. \ref{fig:GB} and \ref{fig:phdepth}, the depth and width of the phase-field depression at the grain boundary depends on the misorientation. As a result, recognition of the small angle grain boundaries becomes more difficult, as they become less visible for $\Delta \theta \rightarrow 0$. The evaluation procedure is illustrated in Fig. \ref{fig:eval}. Figs. \ref{fig:eval}(a) and \ref{fig:eval}(b) show the original orientation field and phase-field distributions. Here, we use  a discretized gray-scale representation of the phase field that has 256 shades. We use the watershed algorithm of MATLAB \cite{ref39} to determine the grain boundary network. This discretization of the gray hue removes a minor ($\sim 10^{-4}$) scattering of the phase field in the vicinity of the grain boundaries that originate  from the fluctuations of the orientation field in the liquid phase. (This scattering would otherwise produce very small grain sizes, when applying the watershed algorithm.) This procedure results in well discernible grain boundaries even in the case of small angle grain boundaries, where the phase-field depression is small [see Fig. \ref{fig:eval}(b)]. One expects that in the experiments recognition of small angle grain boundaries is more difficult. A similar situation occurs here, as the phase-field depression varies with misorientation. In order to understand how the LGSD depends on the recognition of the small angle grain boundaries, we have processed the gray-scale image further by converting it to black and white using a threshold value for phase field, $\phi_{th}$. If $\phi \le \phi_{th}$ the actual pixel is considered as belonging to a grain boundary, and is painted black, the rest is colored white. Increasing $\phi_{th}$, an increasing number of pixels are recognized as belonging to grain boundaries, and an increasing fraction of the shallower depressions representing the small angle grain boundaries are detected [see Figs. \ref{fig:eval}(c)--\ref{fig:eval}(f)]. Evidently, with increasing $\phi_{th}$ an increasing fraction of the simulation box becomes black. The theoretical upper limit to detect any difference using this method is $\phi_{th} > 1 - 1/256$. Once the grain boundary network was computed, the grains that are defined as white areas, whose pixels can all be visited from its other pixels without crossing a grain boundary, using the watershed algorithm. The grains identified so are then divided into size categories, and presented in the form of histograms approximating the respective probability density distribution.

\begin{figure}[b]
 \includegraphics[width=0.99\linewidth]{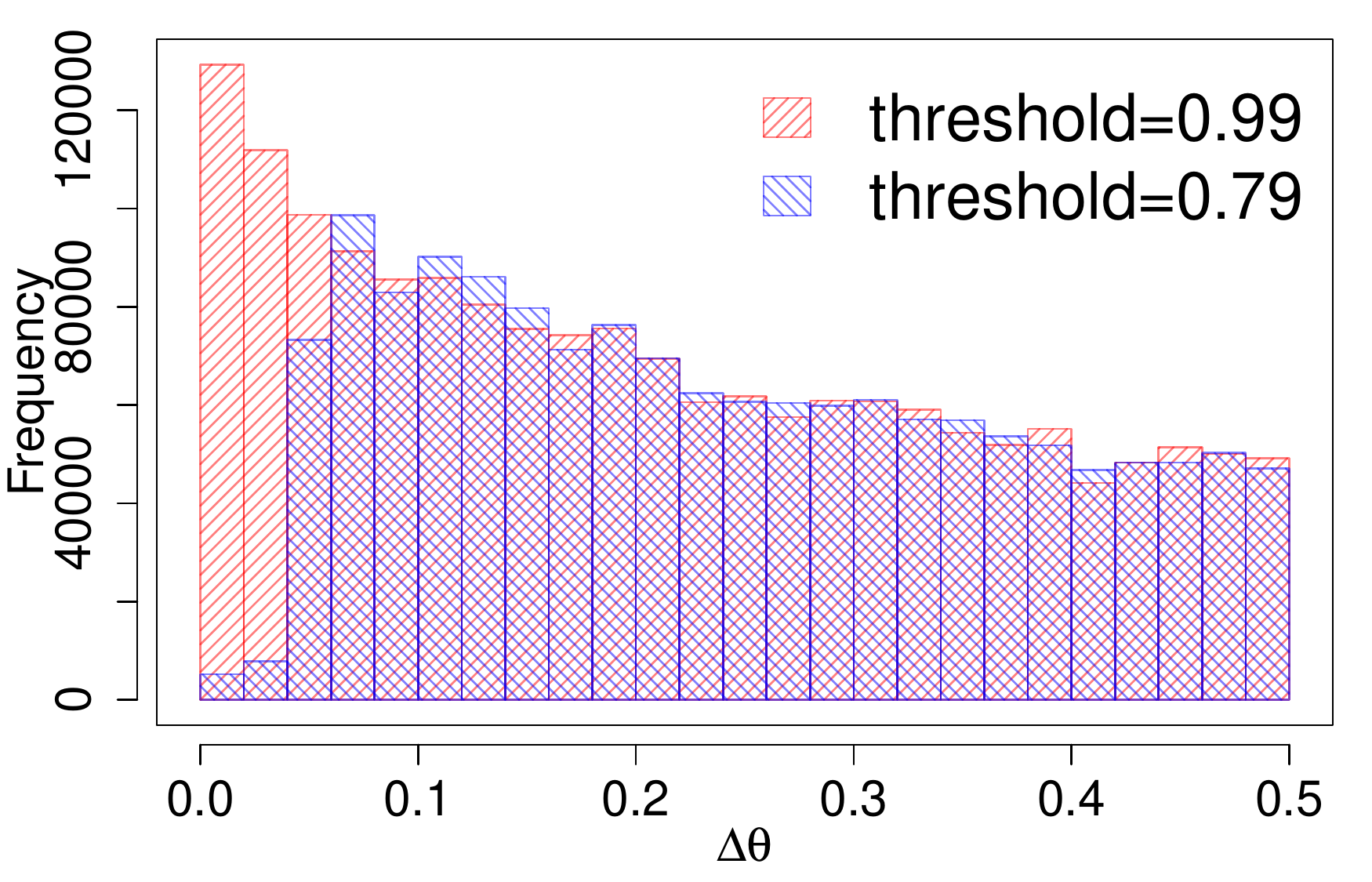}\\  
\caption{\label{fig:pmisori} (color online) Comparison of the histograms characterizing misorientation distributions for thresholds $\phi_{th} = 0.79$ (blue) and $\phi_{th} = 0.99$ (orange) in the case of the HMP simulation at $4 \times 10^6$ time steps. In constructing the histograms 25 equal size misorientation ranges were used. Note the reduced amount of small misorientations in the distribution obtained with $\phi_{th} = 0.79$.}
\end{figure}

\begin{figure}[t]
 (a) \includegraphics[width=0.93\linewidth]{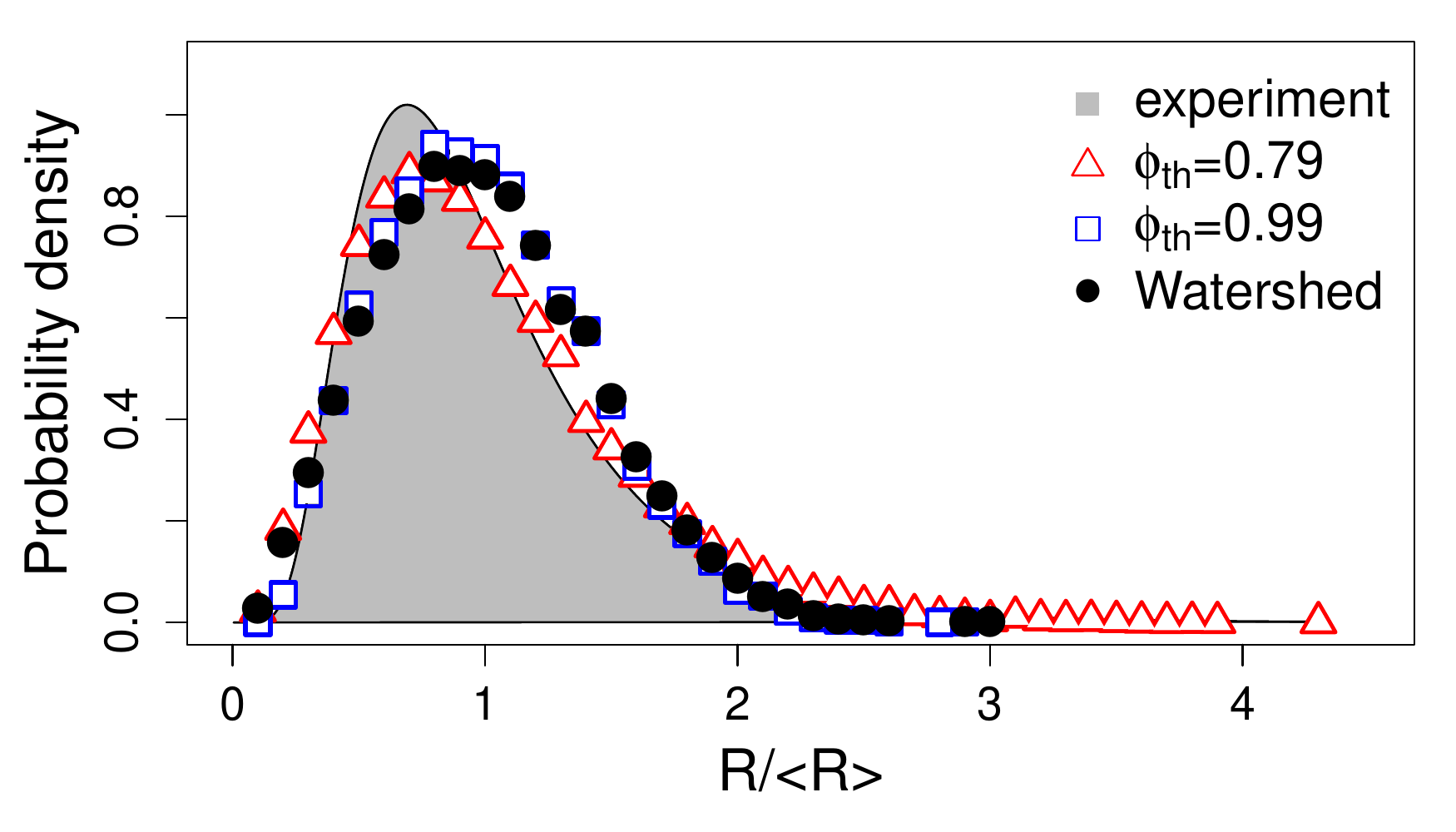}\\  
 (b) \includegraphics[width=0.93\linewidth]{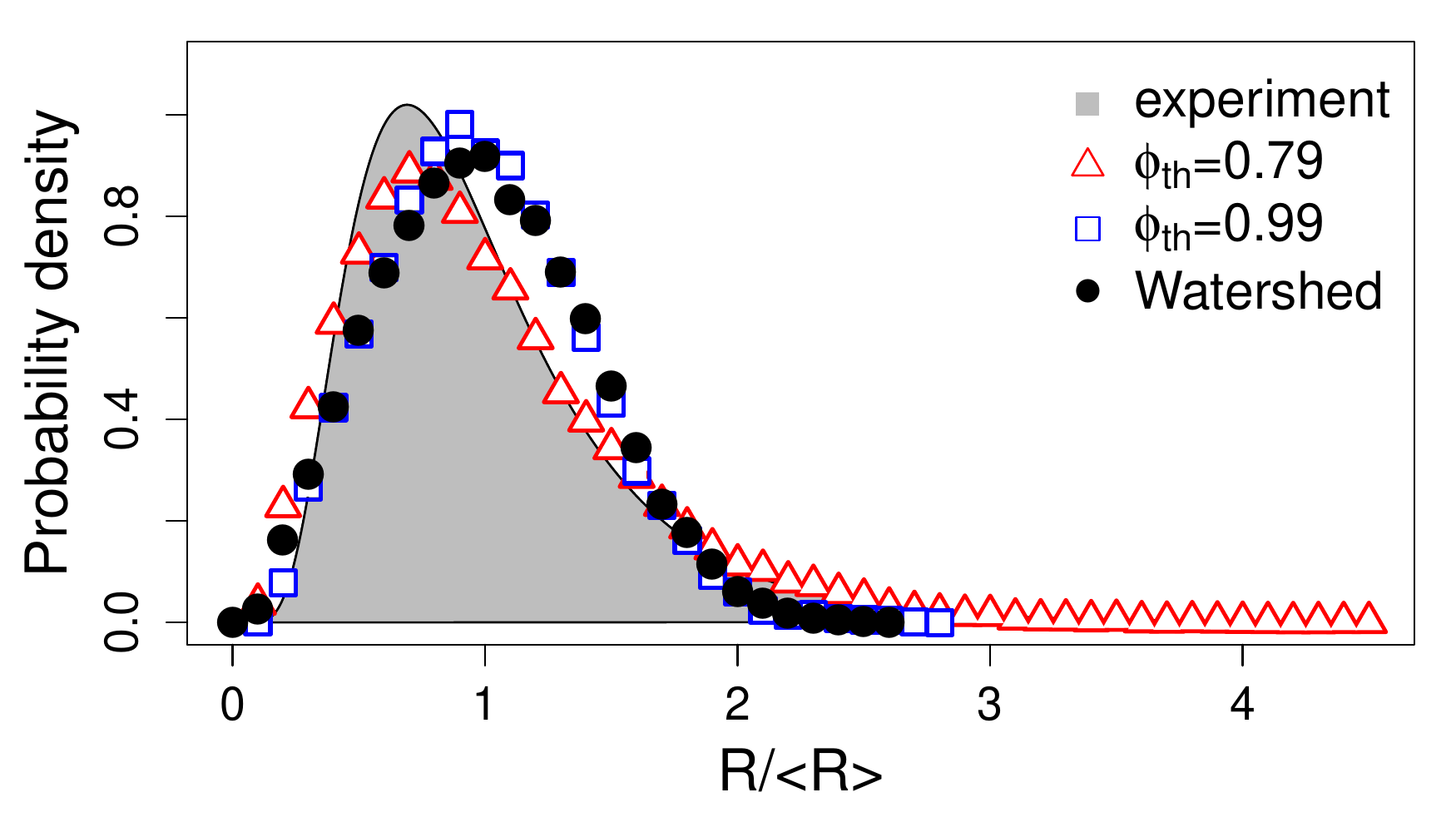}\\   
\caption{\label{fig:goodLGSD} (color online) Limiting grain size distributions for (a) the HMP and (b) KWC simulations. Note the closeness of the  results evaluated using $\phi_{th} = 0.99$ and those evaluated from the grayscale image of the phase field map using the watershed algorithm. It is also remarkable that with the low threshold $\phi_{th} = 0.79$ lognormal type distributions close to the experimental one were obtained.}
\end{figure}

\subsubsection{Varying the threshold}

As one may expect on the basis of Fig. \ref{fig:phdepth}, the LGSD depends on the choice of the threshold $\phi_{th}$. The LGSDs obtained using the threshold values applied in Figs. \ref{fig:eval}(c)--\ref{fig:eval}(f) are compared with each other and the experimental distribution in Fig. \ref{fig:LGSDcomp}.
We have also evaluated the probability distribution of the misorientations at the grain boundaries (see Fig. \ref{fig:pmisori}). A comparison of the distributions corresponding to $\phi_{th} = 0.79$ and $\phi_{th} = 0.99$  clearly indicate that the major difference between the respective misorientation distribution is indeed a lack of small angle misorientations in the former case. 
It appears that with increasing $\phi_{th}$, we see a transition from a distribution falling close to the experimental lognormal distribution towards the generic Mullins-type LGSD the majority of the 2D simulations predict ({\it cf.} LGSD colored blue and red in Fig. \ref{fig:simcomp}). The distributions obtained using $\phi_{th} = 0.99$ are in fact very close to those obtained from the grayscale image of the phase field map using the watershed algorithm (see Fig. \ref{fig:goodLGSD}), implying that with this threshold the majority of the low angle grain boundaries was found.


{\color{blue}These results indicate that the LGSD is critically sensitive to the resolution of the small angle grain boundaries, and raise the possibility that this sensitivity may at least partly be responsible for the deviation between the results of the (fairly coherent) 2D simulations shown in Fig. \ref{fig:simcomp} and other methods that lead to different LGSDs.} We note in this respect that with the exception of approaches based on the OF concept; the simulations shown in Fig. \ref{fig:simcomp} are immune to such errors: e.g., the multi-phase-field models use an individual phase-field for each orientation, so in these models grains can always be distinguished.

\begin{figure}[t]
 (a) \includegraphics[width=0.93\linewidth]{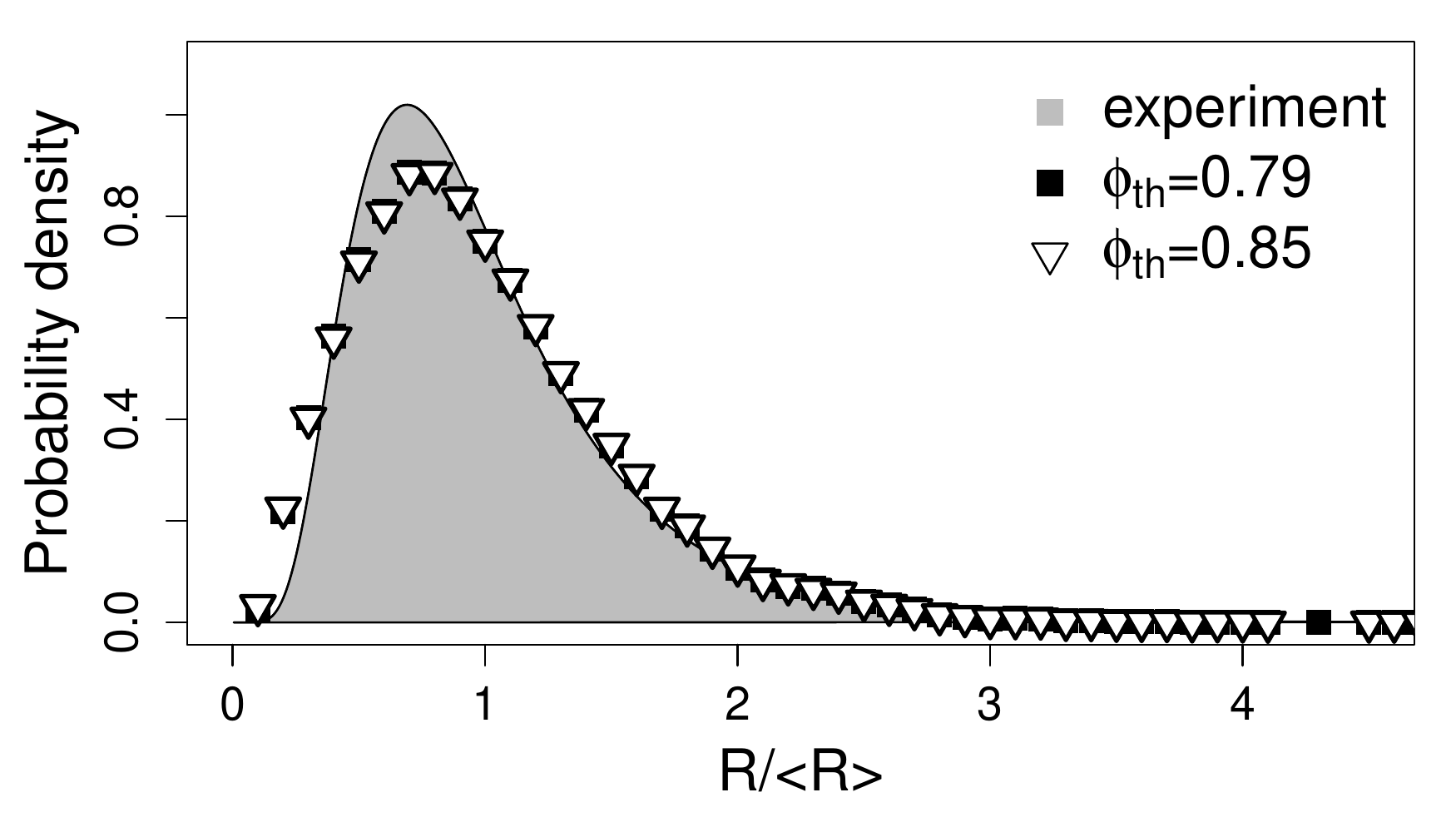}\\  
 (b) \includegraphics[width=0.93\linewidth]{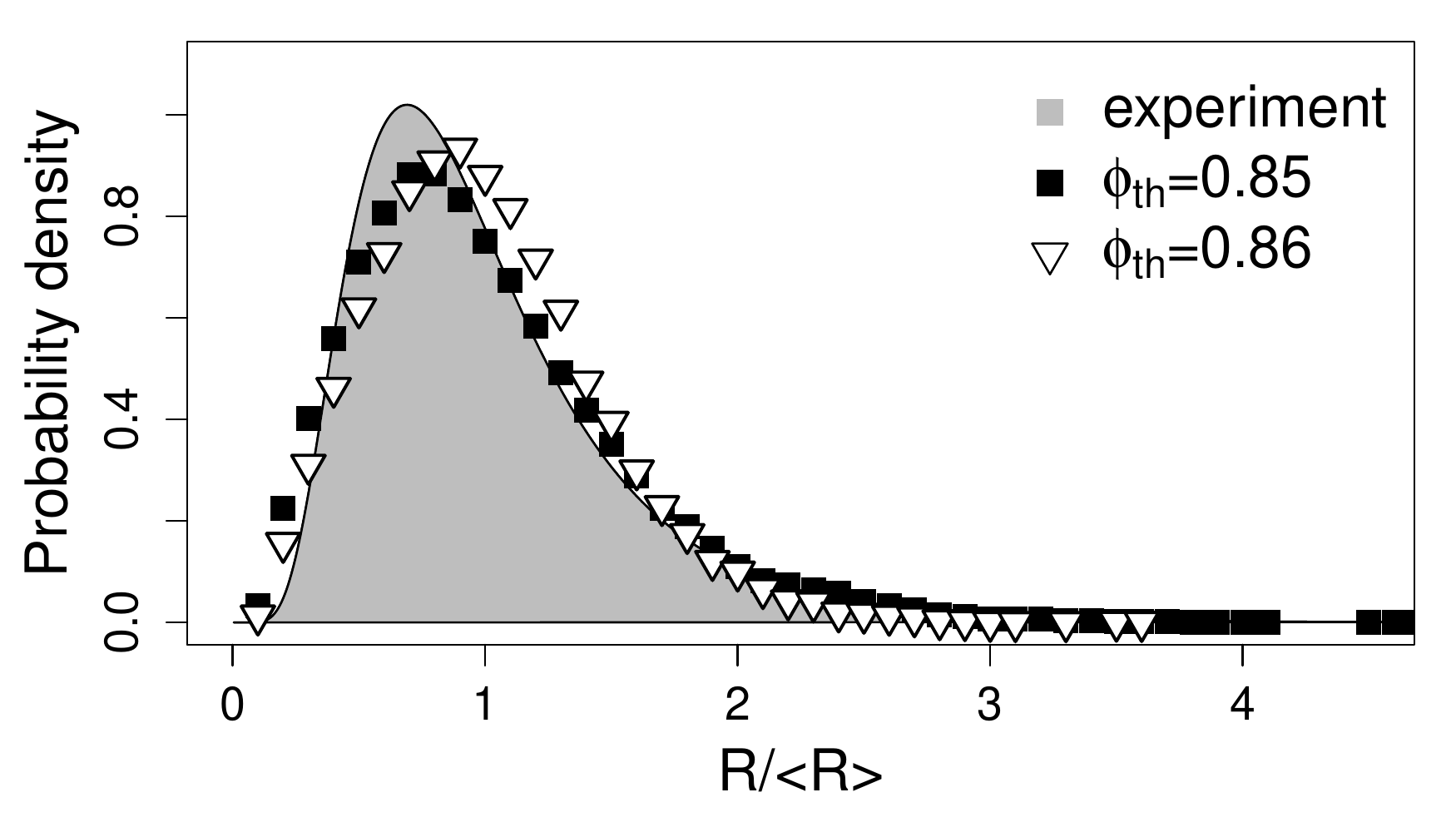}\\   
 (c) \includegraphics[width=0.93\linewidth]{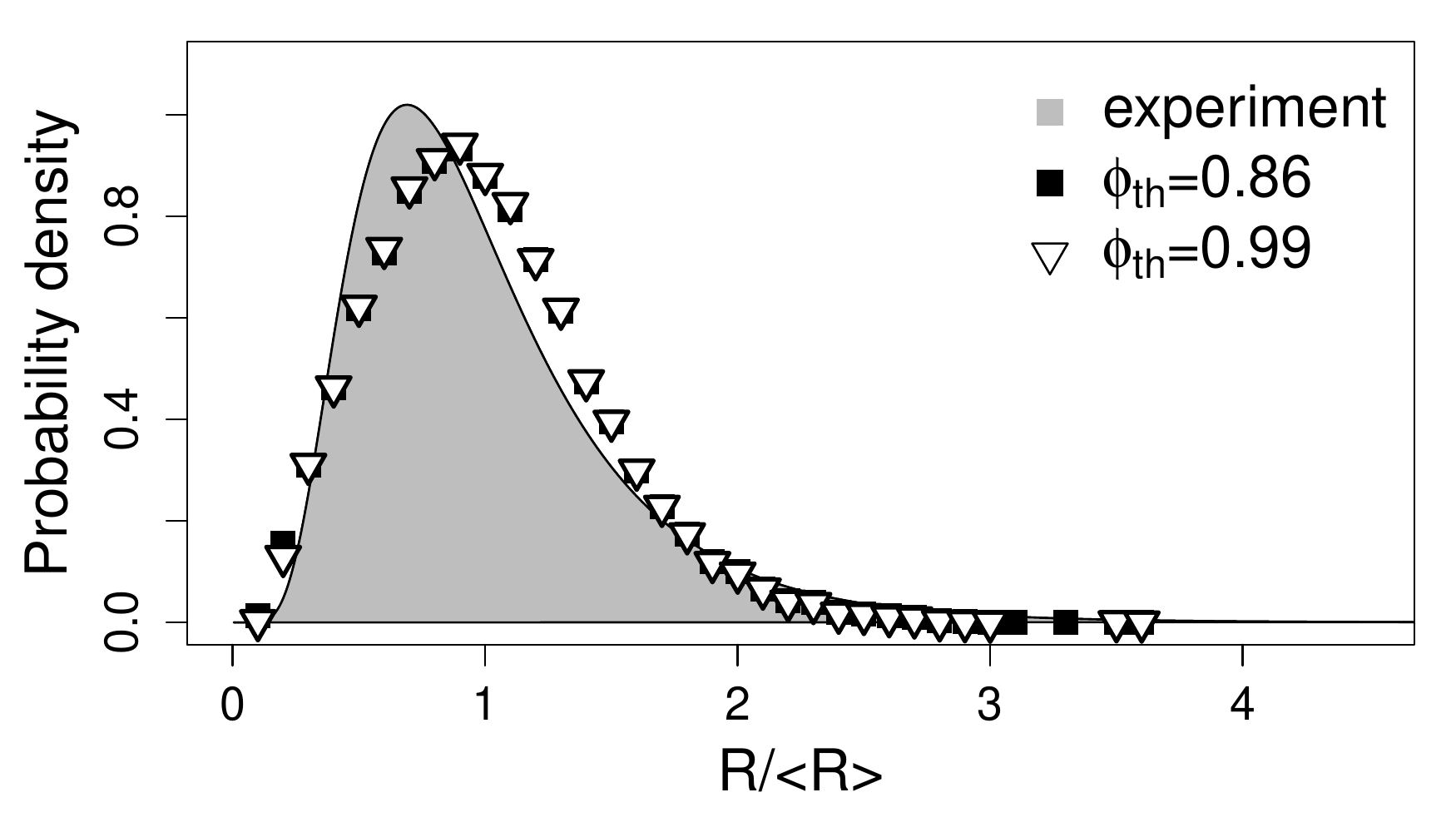}\\  
  
\caption{\label{fig:dLGSDcomp} Limiting grain size distributions for KWC simulations with 30 equidistant orientations, obtained using the threshold values: $\phi_{th} = 0.79$ 0.85, 0.86, and $0.99$. Note that the transition between the lognormal and the Mullins type distributions happens abruptly between $\phi_{th} = 0.85$ and $0.86$. Comparable results were obtained with the HMP model.}
\end{figure}

\begin{figure}[t]
\includegraphics[width=0.99\linewidth]{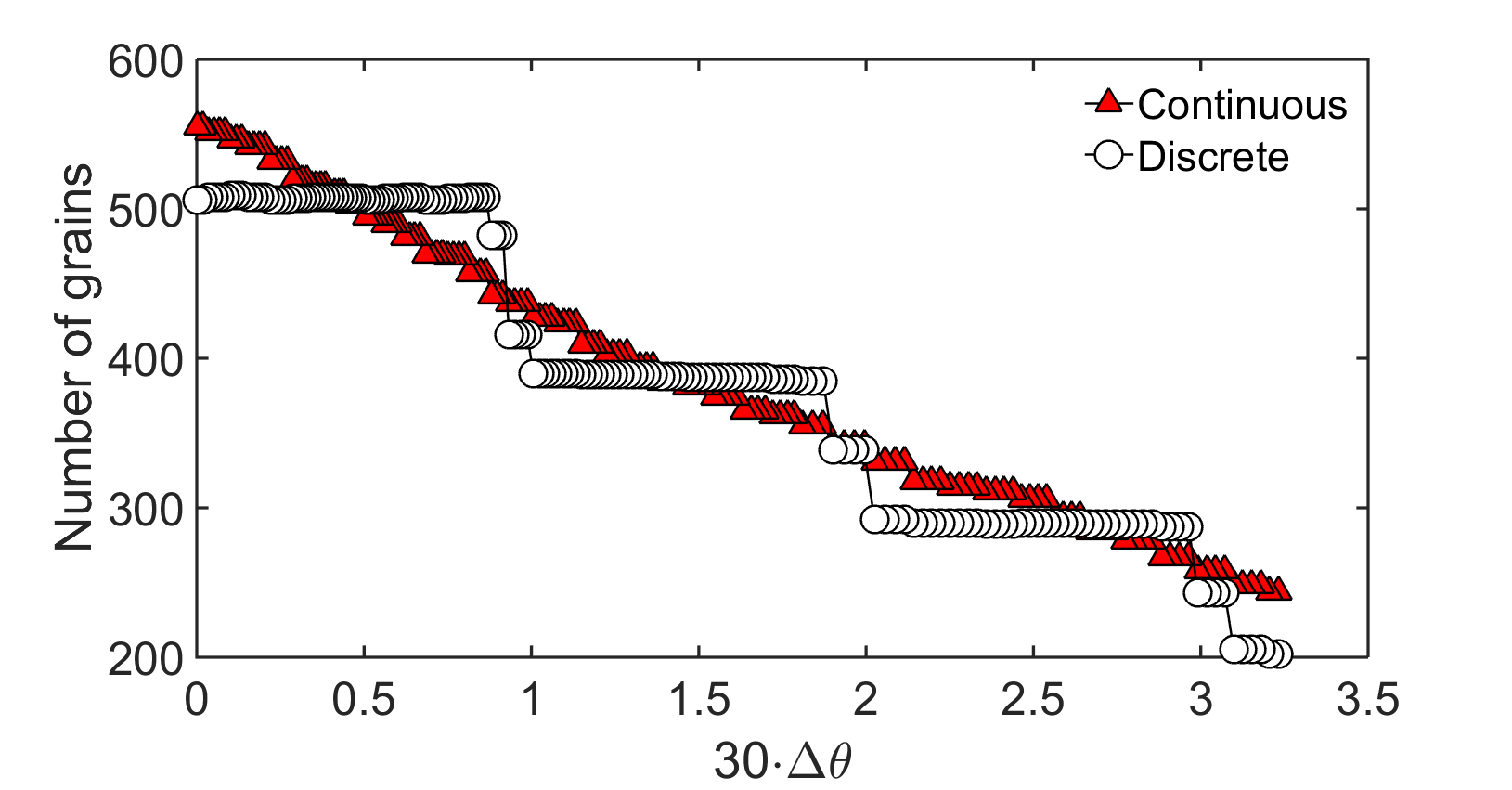}\\    
\caption{\label{fig:stairs} (color online) The number of grains detected as a function of the misorientation threshold in the KWC simulations. Results obtained with continuous (triangles) and discrete (circles) initial orientations of the seeds are shown. Note that in the discrete case, the jumps correspond to multiples of the minimum misorientation $\Delta\theta_\mathrm{min} = 1/30$ defined by the initial set of orientations.}
\end{figure}

\subsubsection{Discrete orientations}   

\begin{figure}[t]
(a)\includegraphics[width=0.93\linewidth]{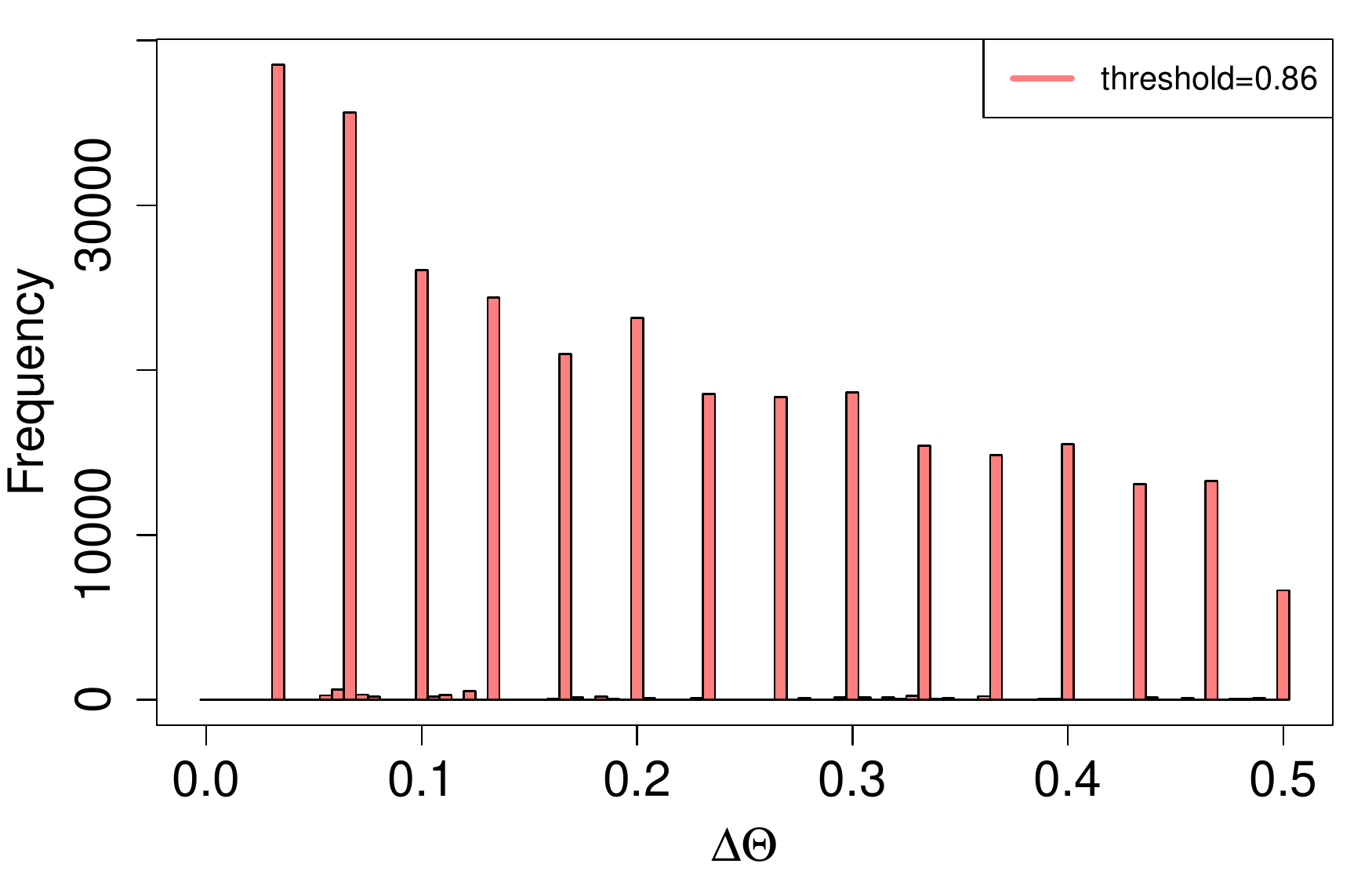}\\    
(b)\includegraphics[width=0.93\linewidth]{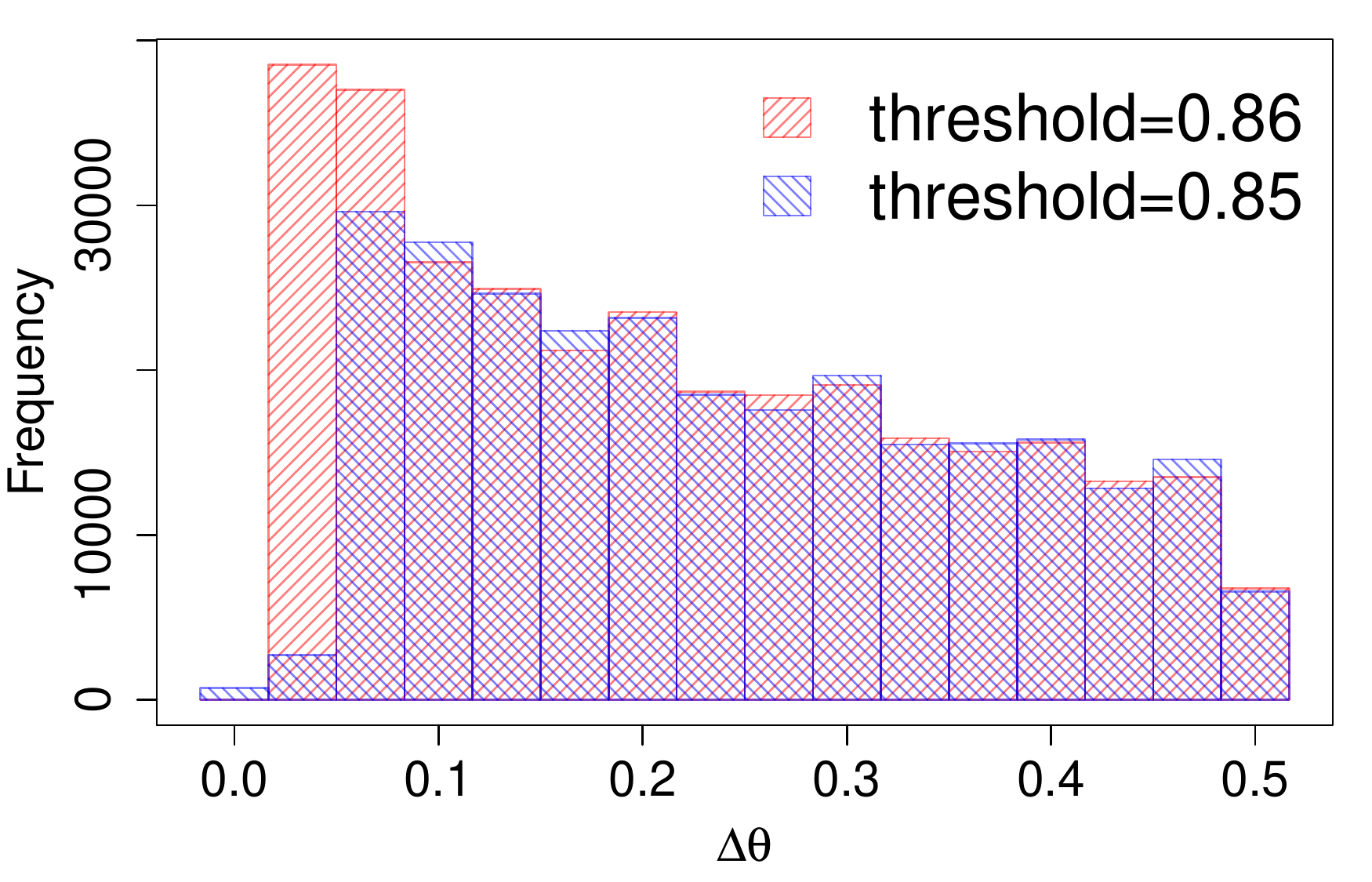}\\    
\caption{\label{fig:ddmisori} (color online) Histograms showing the distribution of the misorientations in the KWC simulations at $2 \times 10^6$ time steps. (a) Histogram obtained for $\phi_{th} = 0.86$ using 90 bins that shows the discrete nature of the orientations. (b) Comparison of histograms obtained for thresholds $\phi_{th} = 0.85$ and 0.86, corresponding to positions slightly below and above the jump at $\Delta \theta = 1/30$ in Fig. \ref{fig:stairs}. (Here 15 bins were used in constructing the histogram.) Note the lack of small angle misorientations in the case of the lower threshold.}
\end{figure}

The MPF simulations are performed usually with a relatively large but finite number of discrete orientations (in early studies about 30 equidistant orientations were regarded as a satisfactory approximation in 2D \cite{ref40}, however, recently $\sim 100$ orientations are considered more appropriate \cite{ref23}). We have tried a similar approach in the orientation field models: we started simulations with initial grains of 30 equidistant orientations of equal probability. Apart from the effect of establishing a continuous transition at the grain boundaries, the initial orientations remained dominant during the process of grain coarsening. This results in grain boundaries with misorientations that are an integer multiples of the minimal misorientation $\Delta \theta_\mathrm{min} = 1/30$. Accordingly, when varying $\phi_{th}$, the change of the LGSD happens stepwise between $\phi_{th} = 0.85$ and $0.86$ for the KWC model, and between $\phi_{th} = 0.81$ and $0.82$ in the case of the HMP model. This stepwise behavior is also visible, when plotting the number of detected grains as a function of the misorientation $\Delta \theta$ that corresponds to the threshold $\phi_{th}$ used in the evaluation process (Fig. \ref{fig:stairs}). For grain orientations  varying continuously, the number of grains also vary continuously, whereas for equidistant discrete orientations the number of grains varies stepwise. The respective misorientation distributions reflect these [see Fig. \ref{fig:ddmisori}(a)]. Furthermore, in the case of the lower threshold ($\phi_{th} = 0.85$), the first peak and part of the second is missing from the misorientation distribution, which explains the abrupt change of the LGSD, and shows again the importance of the amount of small angle grain boundaries in shaping the LGSD [see Fig. \ref{fig:ddmisori}(b)].

\subsubsection{A few additional remarks}

It is  worth mentioning that {\color{blue} although in the OF models the grain boundary energy is misorientation dependent, still it does not depend on the  inclination of the interface, whereas the dependence on inclination is evidently present in the experiments and the PFC simulations. Also,} all the simulations considered herein are strictly two-dimensional. Yet, in principle, it cannot be excluded that as the real experiments are quasi-two-dimensional (the thickness is not fully negligible), they have to be modeled as thin but three-dimensional cases, when addressing the LGSD. {\color{blue}Such studies are, however, out of the scope of the present work.}

{\color{blue} We note furthermore that the present results may have implications regarding the controversy between the lognormal LGSD from experiment and PFC simulations, and the distribution that appears to be a generic (Mullins-type) solution from other 2D simulations: (a) It might be possible that the population of the small angle grain boundaries is indeed lower in the experiments and the PFC model than in the simulations shown in \ref{fig:simcomp}. If so, evaluation of the misorientation distribution may be useful tool for a more complete characterization of the grain coarsening process. (b) Another theoretical possibility is that the resolution of the small angle grain boundaries is less satisfactory in the cases of experiments and PFC simulations. While the TEM experiments (whose results are shown as reference in Figs. \ref{fig:the_vs_exp}, \ref{fig:simcomp}, and \ref{fig:pfc}) are expected to resolve all grain boundaries no matter what the misorientation is \cite{ref41}, other methods may be less successful in detecting the small angle grain boundaries. Concerning the PFC simulations, our experience in analyzing polycrystalline patterns in terms of the complex hexatic order parameter seems to indicate that the evaluation of the numbers of grains may be subject to uncertainties especially when the small angle grain boundaries are defined by only a few dislocations} \cite{ref42}{\color{blue}. Work is underway to clarify these issues further.

Finally, we wish to stress the point that grain coarsening is a complex process, which depends on several factors. Our work shows that minor changes in the detection of low angle grain boundaries can lead to completely different LGSDs, which may hide the effect of other factors. Although our study is limited to the OF models, it indicates that extreme care needs to be taken to ensure the accuracy of the grain size distribution.}

\section{Summary}

Using orientation field based phase-field models we have investigated how the detection of small angle grain boundaries influence the limiting grain size distribution (LGSD), towards which the grain boundary network relaxes at long times. We make the following concluding remarks:

(i) It appears that the 2D computer simulations relying on the orientation field based phase-field models predict LGSDs that are consistent with LGSDs from the Q = 72 state Potts model, Mullins' model, three versions of the multi-phase-field theory, and with results from a numerical surface solver.  These distributions, however, differ significantly from the lognormal LGSD emerging from the experiments and the 2D phase-field crystal model.

(ii) In the orientation field models we have observed that the LGSD is critically sensitive to the detection of the small angle grain boundaries: We introduced an evaluation method in which the variation of a threshold changed the fraction of small angle grain boundaries detected. It has been shown then that considering an increasing fraction of the small angle grain boundaries, the LGSD varies from a lognormal distribution falling close to those emerging from the experiments and the phase-field-crystal model, to a Mullins-type LGSD the other 2D computer simulations predict. The respective changes in the amount of small angle grain boundaries are clearly seen in the misorientation distribution along the grain boundary network.

{\color{blue}
(iii) Further work is needed to clarify whether in the experiments and phase-field-crystal simulation, the population of the small angle grain boundaries is indeed smaller than in other computer simulations, or perhaps other effects are responsible for the observed deviations in the respective LGSDs.}

\begin{acknowledgments} This work has been supported by NKFIH, Hungary under project No. K-115959, by the Hungarian-French Bilateral Scientific and Technological Innovation Fund under Grant No. T\'ET$\_12\_$FR-2-2014-0034; by the ESA MAP/PECS projects “MAGNEPHAS III” (Contract No 40000110756/11/NL/KML) and “GRADECET” (Contract No 40000110759/11/NL/KML), and by the EU FP7 project “EXOMET" (Contract No. NMP-LA-2012-280421, co-funded by ESA). 
\end{acknowledgments}

\section*{Appendix I: Dimensionless form of the orientation field models}
\subsection{Kobayashi-Warren-Carter-model:} 
{\it The free energy functional}:
\begin{align}
F=\int d\mathbf{r} \bigg \{ \frac{\epsilon^{2}T}{2}\left(\nabla\phi\right)^{2}+WTg(\phi)+ [1-p(\phi)]\Delta f + \nonumber \\
+\phi^{4}\left[H_{1}T|\nabla\theta|+H_{2}T(\nabla\theta)^{2}\right] \bigg \},
\end{align}
\noindent where
\begin{center}
\begin{tabular}{ccc}
$\epsilon^{2}=\frac{6\sqrt{2}\gamma_{SL}\delta}{T_{m}}$; &  $W=\frac{6\sqrt{2}\gamma_{SL}}{\delta T_{m}}$; &  $\Delta f=\frac{\Delta H_f}{V_{m}}\left(1-\frac{T}{T_{m}}\right)$; \tabularnewline
$H_{1}=\alpha_{1}\frac{4\gamma_{SL}}{T_{m}}$; & $H_{2}=\alpha_{2}\frac{8\gamma_{SL}\delta}{T_{m}}$. \tabularnewline 
\end{tabular}
\par\end{center}

{\it EOM for the phase-field:} 
\begin{align}
\dot{\phi}=M_{\phi}\bigg \{ \epsilon^{2}T(\nabla^{2}\phi)-WTg'(\phi)+p'(\phi)\Delta f - \nonumber \\
-4\phi^{3}\left(H_{1}T|\nabla\theta|+H_{2}T(\nabla\theta)^{2}\right)\bigg \} 
\end{align}

\noindent measuring the length and time in units $\xi$ and $\tau=\xi^{2}/D_{L}$, respectively, one obtains the following dimensionless equation of motion:
\begin{align}
\dot{\tilde{\phi}}=\tilde{M}_{\phi}\bigg \{ \tilde{\nabla}^{2}\phi-\tilde{W}g'(\phi)+\tilde{\Delta f}p'(\phi)- \nonumber \\
-4\phi^{3}\left(\tilde{H}_{1}|\tilde{\nabla}\theta|+\tilde{H}_{2}(\tilde{\nabla}\theta)^{2}\right) \bigg \} 
\end{align}
\noindent where 
\begin{center}
\begin{tabular}{c}
$\tilde{M}_{\phi}=\frac{M_{\phi}\epsilon^{2}T}{D_{L}}$\tabularnewline
\end{tabular}
\par\end{center}
\begin{center}
\begin{tabular}{c}
$\tilde{W}=\frac{W\xi^{2}}{\epsilon^{2}}=\frac{6\sqrt{2}\gamma_{SL}}{\delta T_{m}}\xi^{2}\frac{T_{m}}{6\sqrt{2}\gamma_{SL}\delta}=\frac{\xi^{2}}{\delta^{2}}$\tabularnewline
\end{tabular}
\par\end{center}
\begin{center}
\begin{tabular}{c}
$\tilde{\Delta f}=\frac{\Delta f\xi^{2}}{\epsilon^{2}T}=\frac{\Delta H_f}{V_{m}}\left(1-\frac{T}{T_{m}}\right)\frac{\xi^{2}}{\epsilon^{2}T}$\tabularnewline
\end{tabular}
\par\end{center}
\begin{center}
\begin{tabular}{c}
$\tilde{H}_{1}=\frac{H_{1}\xi}{\epsilon^{2}}=\alpha_{1}\frac{4\gamma_{SL}}{T_{m}}\frac{\xi T_{m}}{6\sqrt{2}\gamma_{SL}\delta}=\alpha_{1}\frac{\sqrt{2}}{3}\frac{\xi}{\delta}$\tabularnewline
\end{tabular}
\par\end{center}
\begin{center}
\begin{tabular}{c}
$\tilde{H}_{2}=\frac{H_{2}}{\epsilon^{2}}=\alpha_{2}\frac{8\gamma_{SL}\delta}{T_{m}}\frac{T_{m}}{6\sqrt{2}\gamma_{SL}\delta}=\alpha_{2}\frac{2\sqrt{2}}{3}$\tabularnewline
\end{tabular}
\par\end{center}

{\it EOM for the orientation field:}
\begin{equation}
\dot{\theta}=M_{\theta}\nabla\left\{ \phi^{4}\left[H_{1}T\frac{\nabla\theta}{|\nabla\theta|}+2H_{2}T(\nabla\theta)\right]\right\} 
\end{equation}
or in dimensionless form:
\begin{equation}
\tilde{\dot{\theta}}=\tilde{M}_{\theta}\tilde{\nabla}\left\{ \phi^{4}\left[\frac{\tilde{\nabla}\theta}{|\tilde{\nabla}\theta|}+2\tilde{H_{2}^{\theta}}\tilde{\nabla}\theta\right]\right\} 
\end{equation}
\noindent where
$\tilde{M}_{\theta}^{KWC}=M_{\theta}\xi H_1 T/D_{L}$ and $\tilde{H}_{2}^{\theta}=\frac{\tilde{H}_{2}}{\tilde{H_{1}}}=\frac{\alpha_{2}}{\alpha_{1}}\frac{2\delta}{\xi}$.\\

\subsection{The Henry-Mellenthin-Plapp model:}

{\it Free energy functional:}
\begin{align}
F=\int d\mathbf{r} \bigg \{ \frac{\epsilon^{2}T}{2}\left(\nabla\phi\right)^{2}+WTg(\phi)- p(\phi)\Delta f+ \nonumber \\
+ q(\phi)HT(\nabla\theta)^{2}\bigg \},
\end{align}
\noindent where $H=\alpha_{2}\frac{8\gamma_{SL}\delta}{T_{m}}$, whereas $\epsilon^2$, $W$, and $\Delta f$ are the same as for the KWC model.\\

{\it EOM for the phase-field:}
\begin{align}
\dot{\phi}=M_{\phi}\big \{ \epsilon^{2}T(\nabla^{2}\phi)-WTg'(\phi)+ p'(\phi)\Delta f- \nonumber \\
- q'(\phi)HT(\nabla\theta)^{2}\big \}, 
\end{align}

\noindent yielding the following dimensionless equation of motion:

\begin{align}
\tilde{\dot{\phi}}=\tilde{M}_{\phi}\big \{ \tilde{\nabla}^{2}\phi-\tilde{W}g'(\phi)+ \tilde{\Delta f}p'(\phi)- \nonumber \\
- q'(\phi)\tilde{H}(\tilde{\nabla}\theta)^{2}\big \}, 
\end{align}
\noindent where $\tilde{H}=\frac{H}{\epsilon^{2}}=\alpha_{2}\frac{8\gamma_{SL}\delta}{T_{m}}\frac{T_{m}}{6\sqrt{2}\gamma_{SL}\delta}=\alpha_{2}\frac{2\sqrt{2}}{3}$.\\

{\it EOM for the orientation field:}
\begin{equation}
\dot{\theta}=M_{\theta}\nabla\left\{ HTq(\phi)(\nabla\theta)\right\}, 
\end{equation}
and in dimensionless form:
\begin{equation}
\tilde{\dot{\theta}}=\tilde{M}_{\theta}\tilde{\nabla}\left\{ q(\phi)(\tilde{\nabla}\theta)\right\}, 
\end{equation}
while $\tilde{M}_{\theta}^{HMP}=M_{\theta}HT/D_{L}$.

\begin{figure}[t]
 \includegraphics[width=0.99\linewidth]{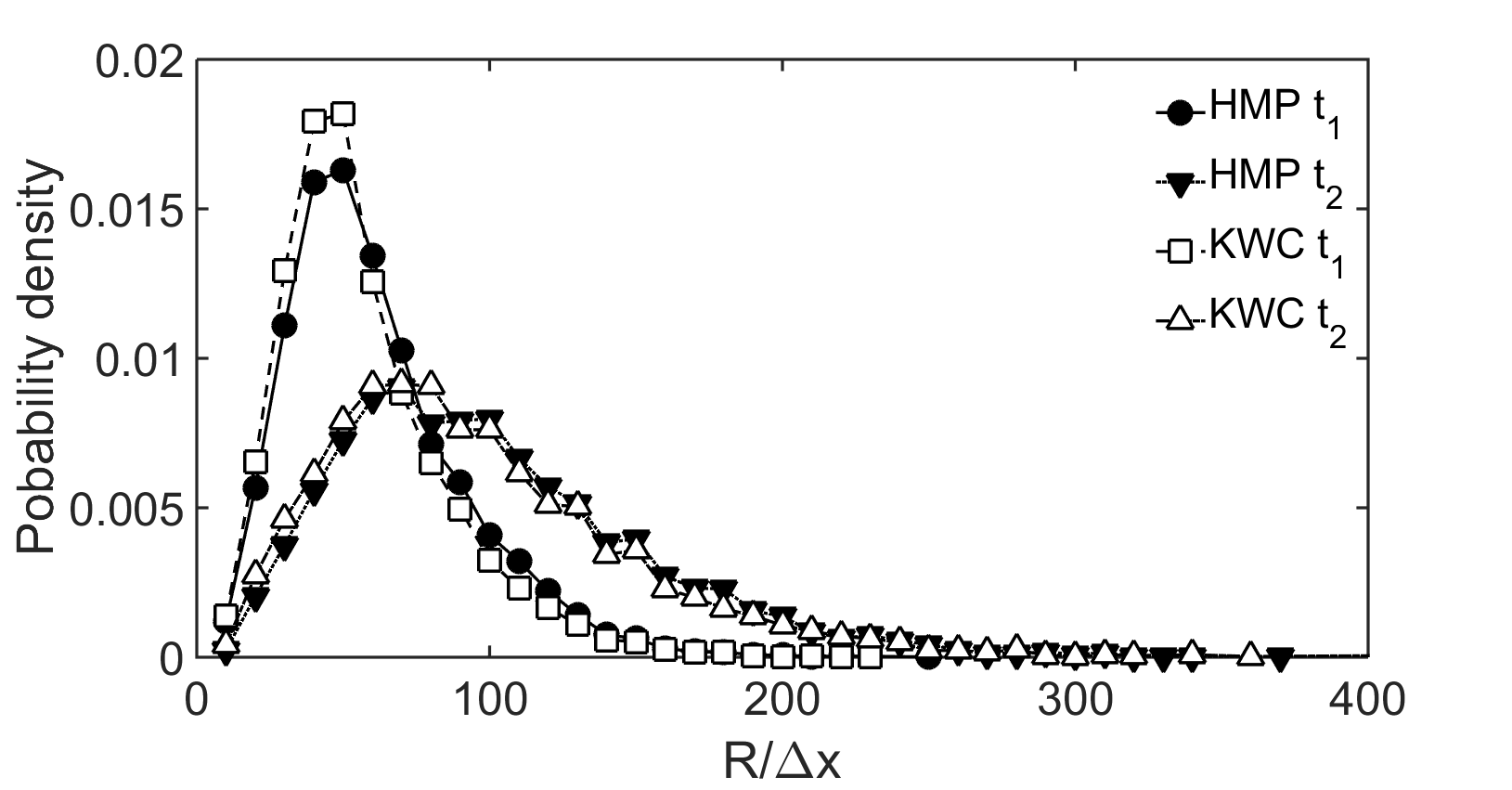}\\  
 \caption{\label{fig:raw} Raw distributions obtained for $\phi_{th} = 0.79$ at times $t_1 = 2 \times 10^6 \Delta t$ and $t_1 = 4 \times 10^6 \Delta t$ for the HMP and KWC models.}
\end{figure}

\begin{figure}[b]
 (a) \includegraphics[width=0.93\linewidth]{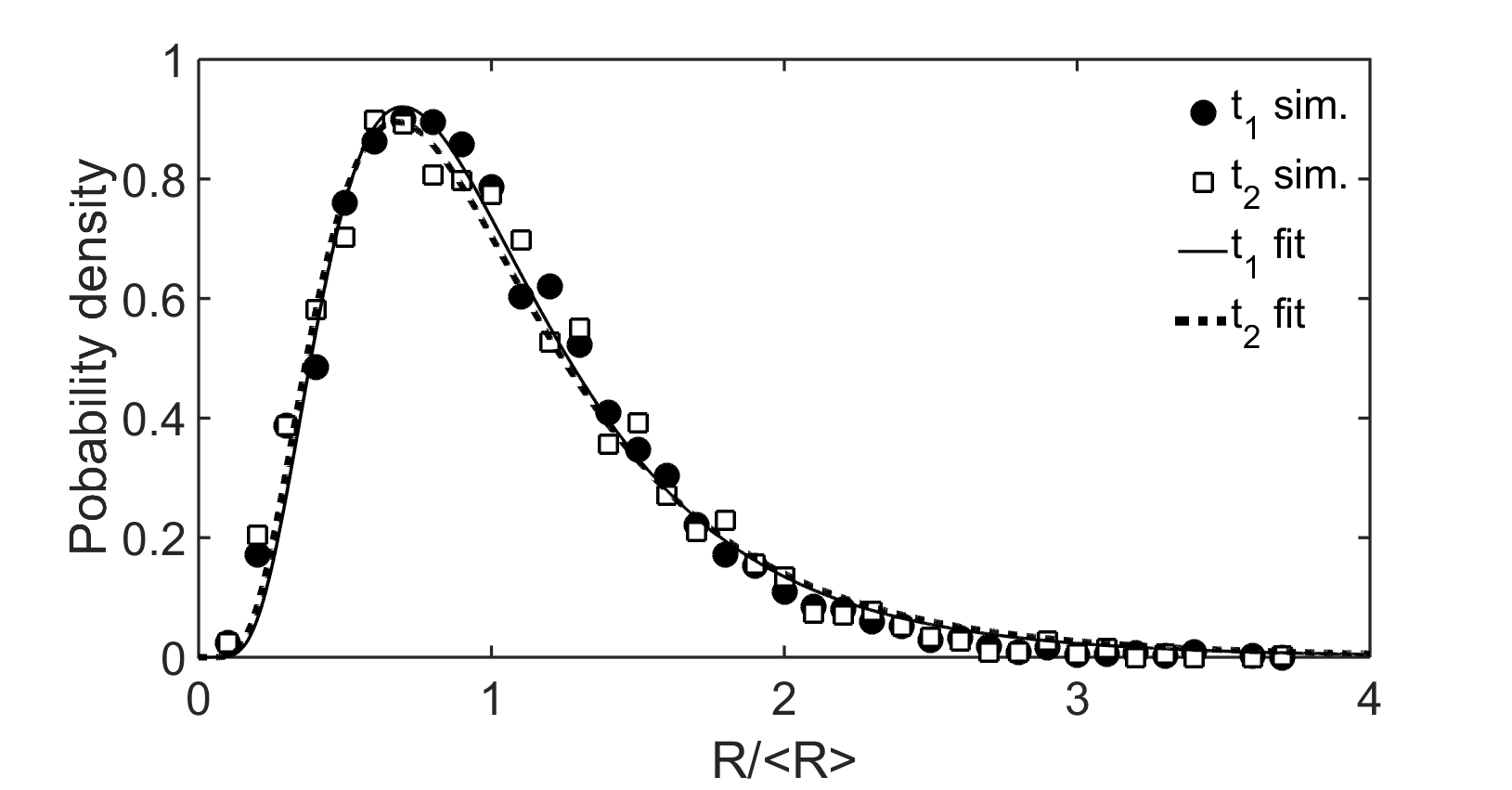}\\  
 (b) \includegraphics[width=0.93\linewidth]{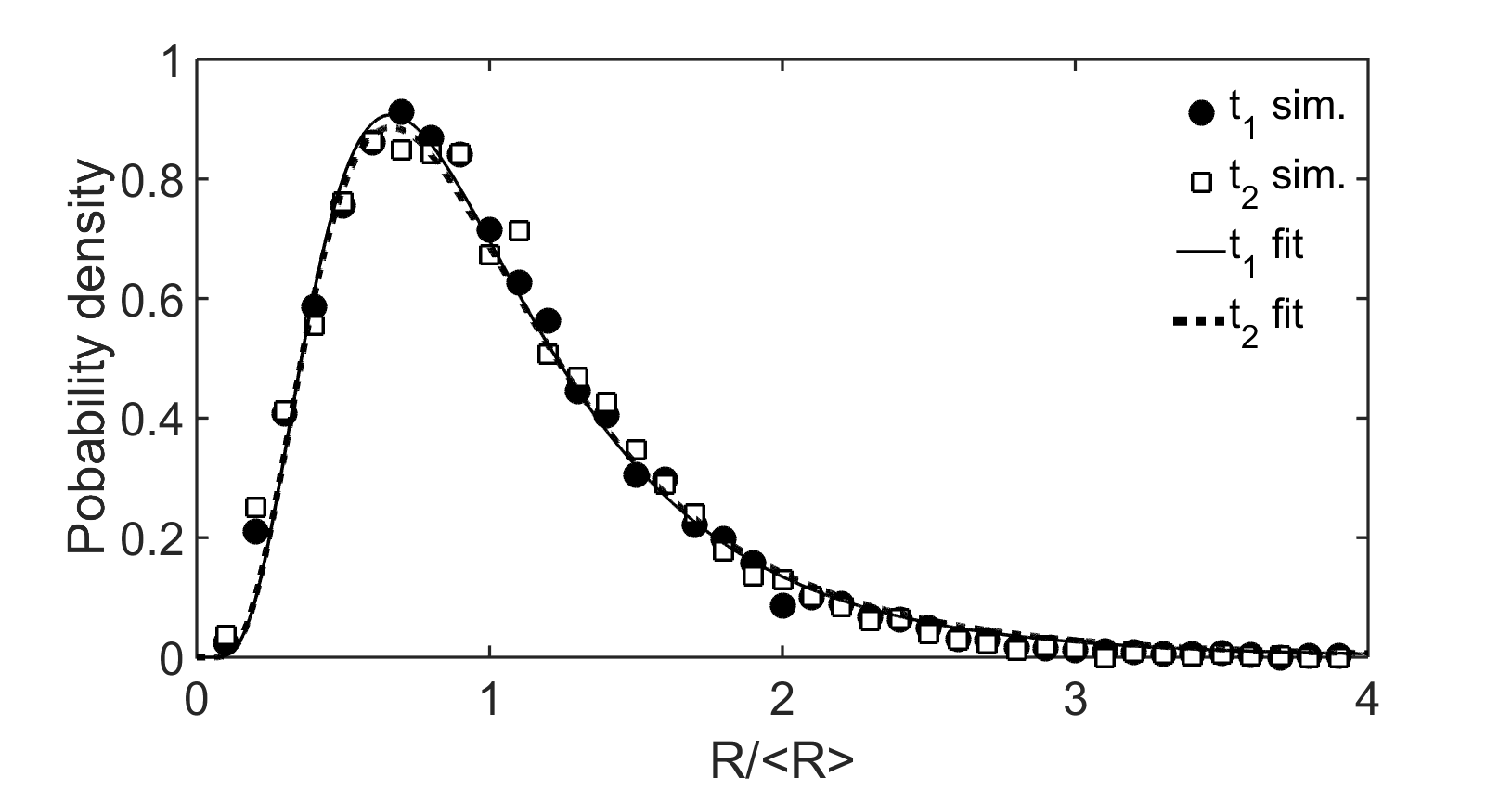}\\   
\caption{\label{fig:reduced} Reduced grain size distributions (symbols) and the fitted lognormal distributions (lines) evaluated at $t_1$ (solid line) and $t_2$  (dashed line) from (a) the HMP and (b) the KWC simulations using the threshold $\phi_{th} = 0.79$ in determining the grain size distribution. For the respective $\langle R \rangle$ data see the text.}
\end{figure}

\begin{figure}[t]
 \includegraphics[width=0.99\linewidth]{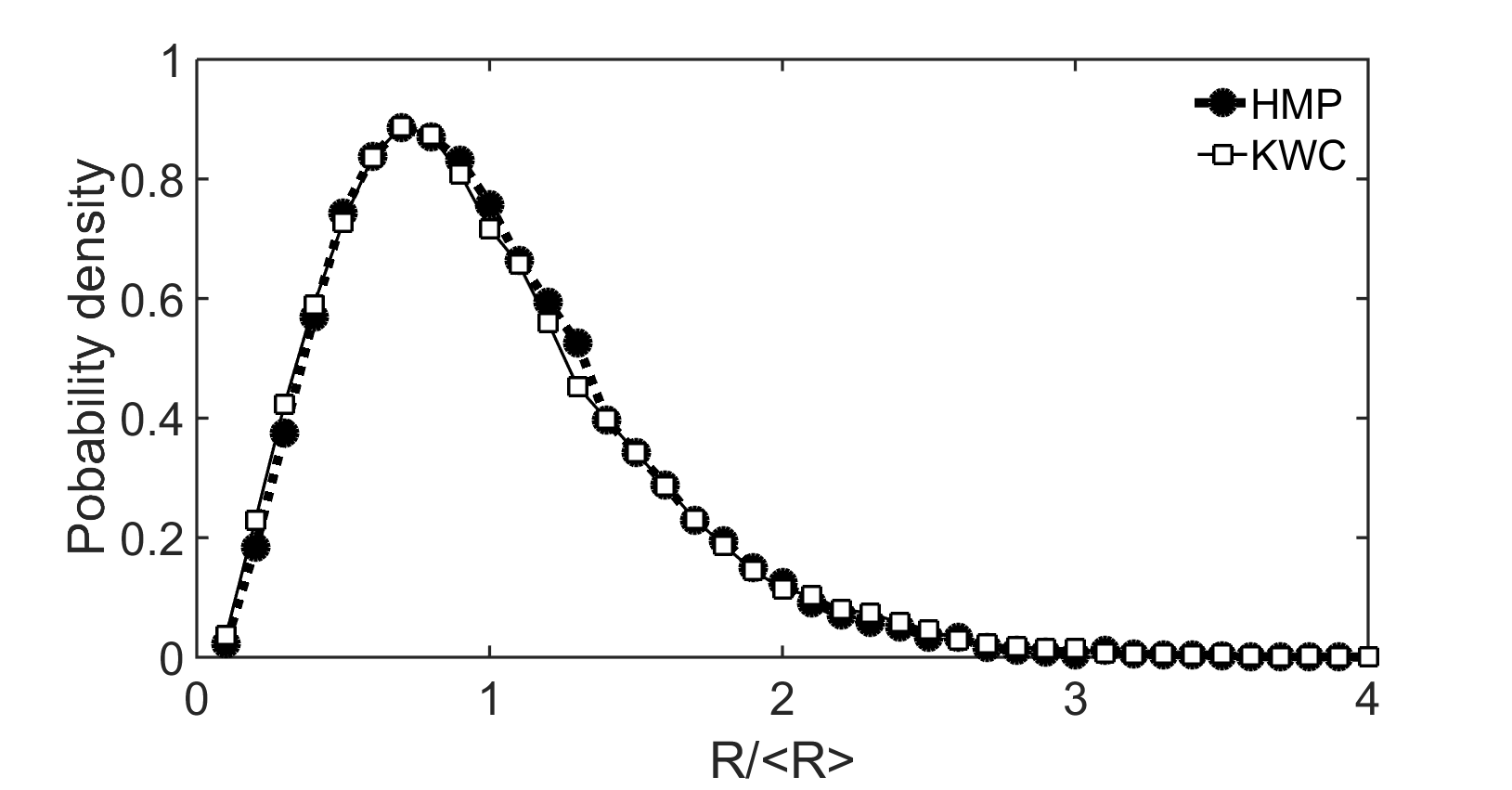}\\  
 \caption{\label{fig:all} The LGSDs obtained by merging 21 reduced distributions evaluated at equidistant instances from $t_1$ and $t_2$ while using a threshold of $\phi_{th} = 0.79$. Note that the orientation field models HMP (filled circles) and KWC (open squares) yield rather similar LGSDs.}
\end{figure}

\begin{figure}[b]
 \includegraphics[width=0.5\linewidth]{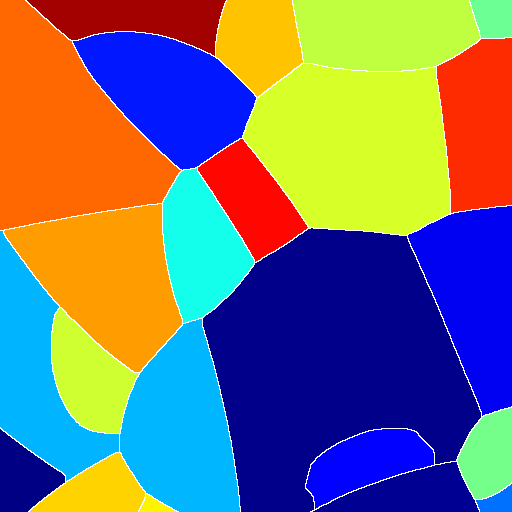}\\  
 \caption{\label{fig:water} (color online) Watershed map corresponding to the phase field map shown in Fig. \ref{fig:eval}(b). The watershed ridges are white, whereas the catchment basins are colored randomly. Note the extra grain detected relative to Fig. \ref{fig:eval}(f), which is hardly visible in the grayscale image.}
\end{figure}

\section*{Appendix II: Reducing the scattering of LGSD}

Under the conditions used herein, practically steady state distribution was achieved at dimensionless time of about $\tilde{t_1} = 2 \times 10^6 \Delta \tilde{t}$ for both the HMP and KWC models. The data representing the state of the system were saved after every $10^5$ time steps until reaching dimensionless time of $\tilde{t_2} = 4 \times 10^6 \Delta \tilde{t}$. The raw distributions corresponding to $\tilde{t_1}$ and $\tilde{t_2}$ are shown in Fig. \ref{fig:raw}. In this period the average size varied from $\langle R \rangle/\Delta x = 61.0$ to $100.7$ for HMP and from $\langle R \rangle/\Delta x = 57.1$ to $96.0$ for KWC. The reduced distributions and the respective fitted (lognormal) distributions are presented in Fig. \ref{fig:reduced}(a) and \ref{fig:reduced}(b) for the HMP and KWC models. The difference of the distributions fitted at the two limiting cases is characterized by the total variational difference $\delta_T = \frac{1}{2} \int_0^{\infty} | p_{t_1}(x) - p_{t_2}(x) | dx$, yielding  $\delta_T = 0.020$ for HMP and 0.011 for KWC, respectively. The LGSDs obtained by merging the respective 21 reduced distributions are reasonably smooth (see Fig. \ref{fig:all}), and enables the detection of small differences between LGSDs that would be hardly perceptible otherwise due to statistical scattering. Here $\phi_{th} = 0.79$ was used. Somewhat larger differences were observed in the case of Weibull fits  to the data from the watershed algorithm ($\delta_T = 0.020$ for HMP and 0.031 for KWC), as there the applied Weibull functions approximate the distributions less accurately.\\

\section*{Appendix III: Evaluation of the misorientation distribution}

In order to characterize the grain boundary network, we have evaluated the distribution of the misorientations weighted with the number of pixels occurring in the watershed map (Fig. \ref{fig:water}). Pixels assigned to the ''catchment basins'' found by the watershed method were associated with the orientations of the respective areas. The number of the pixels in the ''watershed ridge lines'' were used to represent the frequency of the local misorientation. A histogram was made of the latter, an approximation of the probability density distribution of misorientations along the grain boundary network. The misorientation distribution defined so converges towards a limiting distribution at about the same time as the LGSD.



\begin{thebibliography}{99}
\bibitem{ref32} $r(\phi) = \phi^4$ was chosen instead of $r(\phi) = \phi^2$ used in Ref. \cite{ref25} to enhance the depth of the phase-field depression at the grain boundaries, whereas the exponent 3 in the denominator of $q(\phi) = \frac{7\phi^{3}-6\phi^{4}}{(1-\phi)^{3}}$ was chosen instead of 2 in the original paper to avoid the smearing of small angle grain boundaries below a critical misorientation reported for the latter exponent in Ref. \cite{ref26}.
\end{thebibliography}
\end{document}